\documentclass[manuscript]{aastex63}

\usepackage{lineno,hyperref}
\usepackage{amssymb}
\usepackage{graphicx}
\usepackage{array}
\usepackage{sidecap}
\usepackage[section]{placeins}
\usepackage[normalem]{ulem}
\usepackage{nameref}

\hypersetup{linkcolor=red,citecolor=blue,filecolor=cyan,urlcolor=blue}
\received{October 01, 2020}
\revised{April 18, 2021; April 30, 2021}
\accepted{May 4, 2021}
\submitjournal{the Planetary Science Journal}

\shorttitle{Infrared Imaging of SW1}
\shortauthors{Schambeau et al.}
\begin{document}

%\title{Analysis of Spitzer Blue Peak-Up Observations of Centaur 29P/Schwassmann-Wachmann 1 and Refinements to Nucleus Size Estimates: Preparations for Proposed Spacecraft Missions}

\title{Characterization of Thermal Infrared Dust Emission and Refinements to the Nucleus Properties of Centaur 29P/Schwassmann-Wachmann 1}

\correspondingauthor{Charles Schambeau}
\email{charles.schambeau@ucf.edu}

\author[0000-0003-1800-8521]{Charles A. Schambeau}
\affiliation{Florida Space Institute, University of Central Florida \\
12354 Research Parkway, Partnership 1 \\
Orlando, FL 32826, USA}
\affiliation{Department of Physics, University of Central Florida \\
Orlando, FL 32816, USA}

\author[0000-0003-1156-9721]{Yanga R. Fern\'andez}
\affiliation{Department of Physics, University of Central Florida \\
Orlando, FL 32816, USA}
\affiliation{Florida Space Institute, University of Central Florida \\
12354 Research Parkway, Partnership 1 \\
Orlando, FL 32826, USA}

\author[0000-0001-8925-7010]{Nalin H. Samarasinha}
\affiliation{Planetary Science Institute \\
Tucson, AZ 85719, USA}

\author[0000-0003-4659-8653]{M. Womack}
\affiliation{National Science Foundation, Alexandria, VA 22314 USA}
\affiliation{Department of Physics, University of Central Florida \\
Orlando, FL 32816, USA}

\author[0000-0002-8130-0974]{Dominique Bockel{\'e}e-Morvan}
\affiliation{LESIA, Observatoire de Paris, PSL Research University, CNRS, Sorbonne Universit{\'e}, Universit{\'e} de Paris, \\
5 place Jules Janssen, 92195 Meudon, France}

\author[0000-0002-9548-1526]{Carey M. Lisse}
\affiliation{Johns Hopkins University, Applied Physics Laboratory \\
Laurel, MD 20723, USA}

\author[0000-0002-0004-7381]{Laura M. Woodney}
\affiliation{Department of Physics, California State University, San Bernardino\\
San Bernardino, CA 92407, USA}



\begin{abstract}

We present analyses of {\it Spitzer} observations of 29P/Schwassmann-Wachmann 1 using 16 $\mu$m IRS ``blue'' peak-up (PU) and 24 $\mu$m  and 70 $\mu$m MIPS images obtained on UT 2003 November 23 and 24 that characterize the Centaur's large-grain (10-100 $\mu$m) dust coma during a time of non-outbursting ``quiescent" activity. Estimates of $\epsilon f \rho$ for each band (16 $\mu$m (2600 $\pm$ 43 cm), 24 $\mu$m (5800 $\pm$ 63 cm), and 70 $\mu$m (1800 $\pm$ 900 cm)) follow the trend between nucleus size vs. $\epsilon f \rho$ that was observed for the WISE/NEOWISE comet ensemble. A coma model was used to derive a dust production rate in the range of 50-100 kg/s. For the first time, a color temperature map of SW1's coma was constructed using the 16 $\mu$m and 24 $\mu$m imaging data. With peaks at $\sim$ 140K, this map implies that coma water ice grains should be slowly sublimating and producing water gas in the coma. We analyzed the persistent 24 $\mu$m ``wing" (a curved southwestern coma) feature at 352,000 km (90$''$) from the nucleus attributed by \cite{stansberry_2004} to nucleus rotation and instead propose that it is largely created by solar radiation pressure and gravity acting on micron sized grains. We performed coma removal to the 16 $\mu$m PU image in order to refine the nucleus' emitted thermal flux. A new application of the Near Earth Asteroid Thermal Model (NEATM; \cite{harris_1998}) at five wavelengths (5.730 $\mu$m, 7.873 $\mu$m, 15.80 $\mu$m, 23.68 $\mu$m, and 71.42 $\mu$m) was then used to refine SW1's effective radius measurement to $R = 32.3 \pm 3.1$ km and infrared beaming parameter to $\eta = 1.1 \pm 0.2$, respectively.

% We estimate a dust mass fraction for water ice grains in the range of 25-50\% based on the peak temperature from the color map, with smaller grains having a higher ice content. Analysis of coma morphology detected in the 16 $\mu$m and 24 $\mu$m images may indicate a complex icy grain dust coma undergoing fragmentation of grains larger than 24 $\mu$m down to smaller, more stable micron sized grains with higher internal strength.

\end{abstract}

\keywords{general, centaurs --- 
individual, 29P/Schwassmann-Wachmann 1 --- {\it Spitzer Space Telescope} --- infrared observations}


\section{Introduction}

29P/Schwassmann-Wachmann 1 (SW1) is a continuously active Centaur at the inner cusp of the Centaur-to-Jupiter-Family transition region and presents a rare opportunity to investigate activity drivers and ongoing material processing that occurs in a region too cold for vigorous water-ice sublimation. Recent dynamical simulations have shown that its current nearly-circular trans-Jovian orbit (eccentricity, semi-major axis and perihelion respectively: $e$ = 0.04, $a$ = 6.03 au and $q$ = 5.77 au)\footnote{Minor Planet Circular (MPC) 111773.} is typical for Centaurs in a short-lived transitional ``gateway'' from the outer solar system to the Jupiter-family comets (JFCs) population \citep{sarid_2019}. Interestingly, despite SW1's modest variation in energy input from the Sun, it frequently undergoes major outbursts superimposed on its normally-present background, or ``quiescent'' coma \citep{whipple_1980, larson_1980, jewitt_1990_sw1, 2010MNRAS.409.1682T, 2013Icar..225..111K, hosek_2013, miles_2016, schambeau_2017, schambeau_2019}. Additionally, the CO-production rate during periods of quiescent activity is more similar to long-period comets at similar heliocentric distances than JFCs \citep{bauer_2015, kacper_2017, womack_2017, bockelee_2021}, and its dust outbursts may be uncorrelated with large fluctuations of its CO outgassing rate \citep{wierzchos_2020}. Thus questions naturally arise as to what activity drivers explain its enigmatic activity, and do all JFCs experience a period of similar behaviors while they are in the gateway region? Are SW1's activity behaviors reflective of outer solar system materials being thermally activated in the gateway, after a long period of cryogenic storage? Or, are they an intrinsic property to it alone?

In 2015, we reported a new analysis of 2003 November {\it Spitzer} Infrared Array Camera (IRAC) 5.8 $\mu$m \& 8.0 $\mu$m and Multiband Imaging Photometer (MIPS) 24 $\mu$m \& 70 $\mu$m imaging, originally published by \cite{stansberry_2004}. Using a new Spitzer data pipeline and intensive image processing techniques, the 2015 paper presented a new nucleus radius, beaming parameter, and infrared geometric albedo of SW1 \citep{schambeau_2015}. Subsequently, we determined that the {\it Spitzer} ``blue" (i.e. at 16 $\mu$m) images obtained in the 2003 dataset have sufficient coma detected for its analysis, modeling, and removal, and thus, they can provide new physical insights and constraints to SW1 models. 

Here, for the first time, we present the {\it Spitzer} 16 $\mu$m images, and analyze them in the context of the  5.8 $\mu$m, 8.0 $\mu$m, 24 $\mu$m, and 70 $\mu$m data. We describe relevant observational details of the UT 2003 Nov. epoch in Section \ref{sec:observations}. In Section \ref{sec:image_analysis}, we present characterization of thermal infrared emission using the 16 $\mu$m, 24 $\mu$m, and 70 $\mu$m imaging data through coma morphology analysis, estimates of the $\epsilon f \rho$ parameter, coma modeling of the dust grain size distribution and dust-production rates for micron sized and larger grains, and derivation of a coma color temperature map. In Section \ref{sec:neatm} we present analysis of these images to provide a fifth nucleus photometry measurement at 16 $\mu$m. Using the five spectral flux density measurements of the nucleus, we implemented a NEATM \citep{harris_1998} to derive a new measurement of the nucleus' effective size and infrared beaming parameter ($\eta$; a proxy for nucleus surface thermal inertia and/or surface roughness).  In Section \ref{sec:conclusion} we summarize our results and implications for SW1's nucleus, quiescent large grain coma, and activity state. 

\section{Observations} \label{sec:observations}
This work analyzes the {\it Spitzer} imaging data obtained with the 16 $\mu$m IRS blue PU and 24 $\mu$m and 70 $\mu$m MIPS instruments. Here we address the observational details of the 16 $\mu$m data, and direct readers to our earlier work, \cite{schambeau_2015}, for information about the 24 $\mu$m and 70 $\mu$m images.

\begin{figure}
\gridline{
		\fig{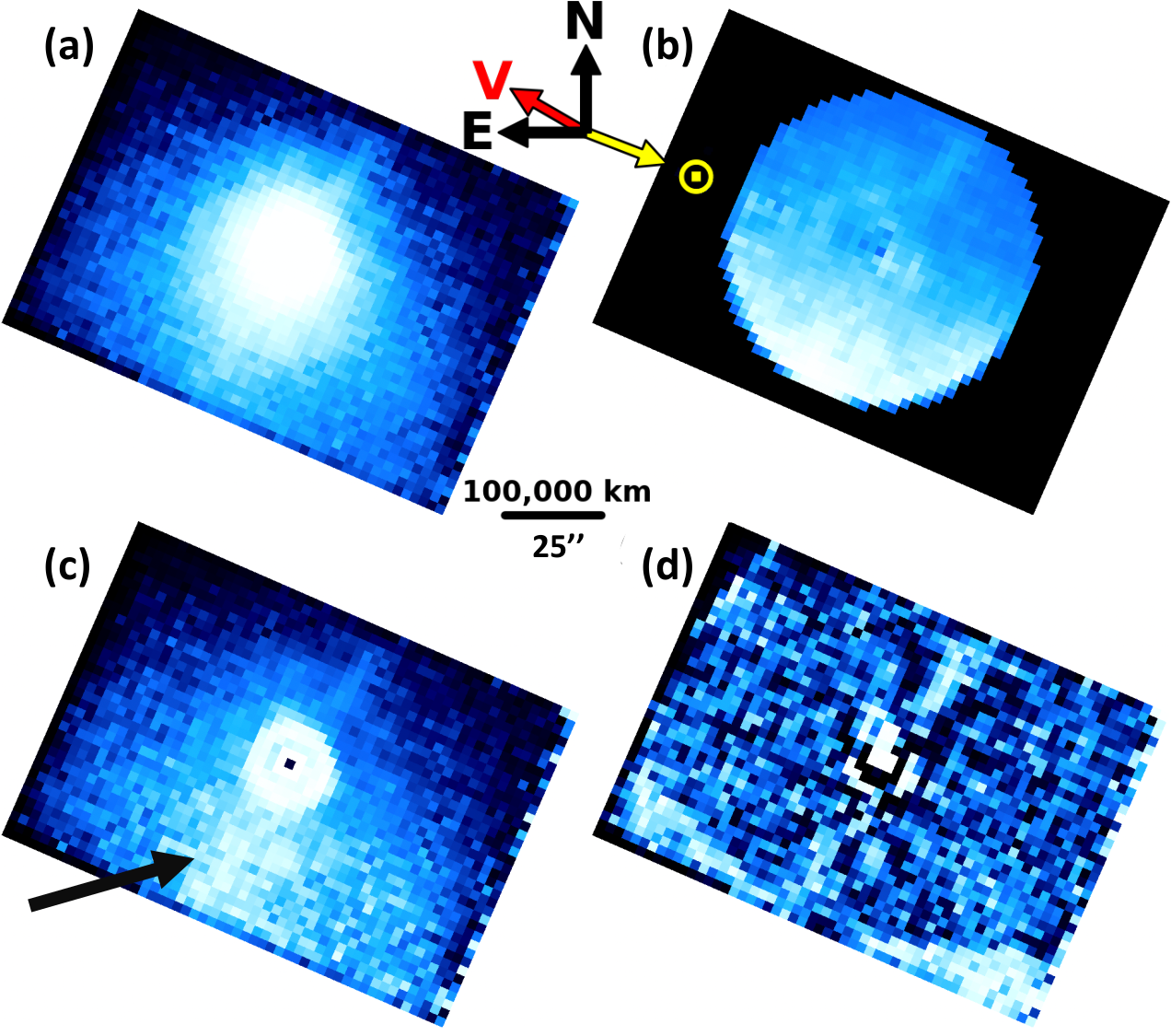}{0.8\textwidth}{}
          }
\caption{One of the {\it Spitzer} 16 $\mu$m blue PU images (with color scale black to blue to white indicating increasing surface brightness): (a) original image, (b) division by azimuthal average, (c) 1/$\rho$ profile removal, and (d) rotational shift differencing of 18$^{\circ}$ \citep{larson_1984, samarasinha_larson_2014}. Equatorial north and east are indicated. The skyplane projected directions for the Sun and SW1's heliocentric velocity vector are indicated by the yellow and red arrows, respectively. A black arrow on panel (c) indicates the slight coma enhancement in the south-southeast direction that is the south-southeast end of the 1 to 7 o'clock linear feature. \label{fig:images}}
\end{figure}

During the {\it Spitzer} in-orbit checkout and science verification phase \citep{werner_2004_spitzer} SW1 was observed with the InfraRed Spectrograph (IRS; AORKEY: 6068992; \cite{houck_2004_IRS}). Shortly before the IRS observations, blue-channel PU images were acquired in order to center SW1's position on the detector's ``sweet spot" (the detector pixel location of the target's centroid peak enabling optimal alignment and centering for the IRS slit). The blue PU channel of IRS's Si:As array detector has dimensions of 44 $\times$ 31 pixels, an effective monochromatic wavelength equivalent to 15.8 $\mu$m, and an effective pixel scale of 1$''$.85/pixel in detector X direction and 1$''$.82/pixel in detector Y direction. A total of six independent blue PU images were acquired: three images with SW1's peak located on the center of the detector and three on the detector's sweet spot, approximately 3 pixels away from the center of the array. Level 1 basic calibrated images were downloaded from the Spitzer Heritage Archive (SHA). An example image of SW1 located on the sweet spot is shown in Figure \ref{fig:images} along with enhanced images to highlight the coma's morphology \citep{larson_1984, samarasinha_larson_2014}. Table \ref{tab:geometry} provides a summary of the observational circumstances. The coma is slightly enhanced in the south-southeast direction and has a similar morphology to that seen in the MIPS 24 $\mu$m images, suggesting that the same particles are being measured in both bandpasses. Overall, aside from the slight increase in dust emission on the south-southeast side of the coma, as indicated by the division of an azimuthal average enhanced image (Figure \ref{fig:images}(b)), the coma is lacking any defining coma morphology. A faint linear feature can be seen from approximately the 1 o'clock to 7 o'clock positions.

\begin{deluxetable*}{ l c c c  c c }
\tablenum{1}
\label{tab:geometry}
\tablecaption{Observations and Geometry Summary for UT 2003 November 23}
\tablewidth{0pt}

\startdata \\
Parameter										&	&	&	&	Value 					\\ \hline
Observations [Start:Stop]							&	&	&	&	[(07:15:32.960):(07:17:13.409)] \\
Exposure Time Per Image							&	&	&	&	9.44 s 					\\
Heliocentric Distance to SW1 ($R_H$)						&	&	&	& 	5.73 au 					\\
Spitzer-SW1 Distance ($\Delta$)					&	&	&	& 	5.54 au 					\\
Solar Phase Angle of SW1 ($\alpha$)							&	&	&	&	10.0$^{\circ}$ 				\\
True Anomaly of SW1									&	&	&	&	342.8$^{\circ}$ 				\\
Position Angle of Skyplane Projected Sun Direction$^{\textrm{a}}$				&	&	&	&	248.2$^{\circ}$				\\
Position Angle of Skyplane Projected Heliocentric Velocity Vector$^{\textrm{a}}$	&	&	&	&	59.9$^{\circ}$				\\
\enddata

\tablecomments{ $^{\textrm{a}}$ The position angle is measured counter clockwise from north through east.}

\end{deluxetable*}

For reference, the filter bandpasses of the 16, 24, and 70 $\mu$m images are respectively: 13.3$-$18.7 $\mu$m, 20.8$-$26.1 $\mu$m, and 60.9$-$80.6 $\mu$m.

%\FloatBarrier
\section{Image Analysis and Discussion}

\subsection{Thermal Infrared Coma Analysis}\label{sec:image_analysis}

Thermal infrared imaging of cometary dust comae allows for preferential probing of grain sizes on the order of microns and larger, such as those recorded with the IRS PU and MIPS, because smaller grains with 2$\pi a/\lambda <$ 1, where $a$ is the grain radius, are inefficient emitters in the infrared \cite[see][]{hanner_1994_apj, lisse_1998, lisse_2004}. Our {\it Spitzer} 16 $\mu$m, 24 $\mu$m, and 70 $\mu$m images were analysed to characterize the continuum emission created by $\mu$m-sized and larger grains in SW1's quiescent dust coma. We note that micron and sub-micron sized grains also contain silicate emission bands between $\sim 8-13$ $\mu$m and  at $\sim$ 20 $\mu$m, which probably contribute a few percent to the flux in the 24 $\mu$m images \citep[see][Figures 13 and 14]{schambeau_2015}. However, a detailed analysis of these emission features and their relatively minor impacts on the 24 $\mu$m imaging is beyond the scope of our current work.
 
 In this section we take advantage of these thermal infrared images in combination with {\it Spitzer}'s stable and well characterized point spread function (PSF) in order to accurately isolate SW1's dust coma flux contributions in each image. We assumed that the dominant grain sizes contributing to the detected flux in each band were approximately the size of their effective monochromatic bandpass wavelengths: 15.80 $\mu$m, 23.68 $\mu$m, and 71.42 $\mu$m (as used by e.g., \cite{bauer_2015, bauer_2017}).

To aid in the analysis of comae morphology, it is useful to reference an idealized ``canonical" coma, containing an isotropic and steady state emission of dust grains from the nucleus, with negligible dust grain fragmentation and solar radiation pressure. This canonical coma has a  surface brightness profile following a $1/\rho$ behavior (where $\rho$ is the skyplane projected conetocentric distance from the nucleus's position), and is assumed in the derivation of the often used $A f \rho$ and $\epsilon f \rho$ parameters \citep{ahearn_1984, lisse_dust_2002, kelley_2013} that are described in more detail in Section \ref{sec:efrho}. In practice the assumptions used to derive $\epsilon f \rho$ break down for real comae, but its calculation provides a first order estimate of comae dust production behaviors. SW1 experienced quiescent activity for at least two months surrounding the UT 2003 Nov. epoch of {\it Spitzer} observations, based on Minor Planet Center (MPC) reported magnitude measurements\footnote{Minor Planet Circulars: 49762, 49871, 49872, 49873, 50347, 50348.}, so the canonical coma assumption is reasonable for these observations.

\subsubsection{16 $\mu$m and 24 $\mu$m Coma Morphology} \label{sec:coma_morphology}

\begin{figure}
\gridline{
		\fig{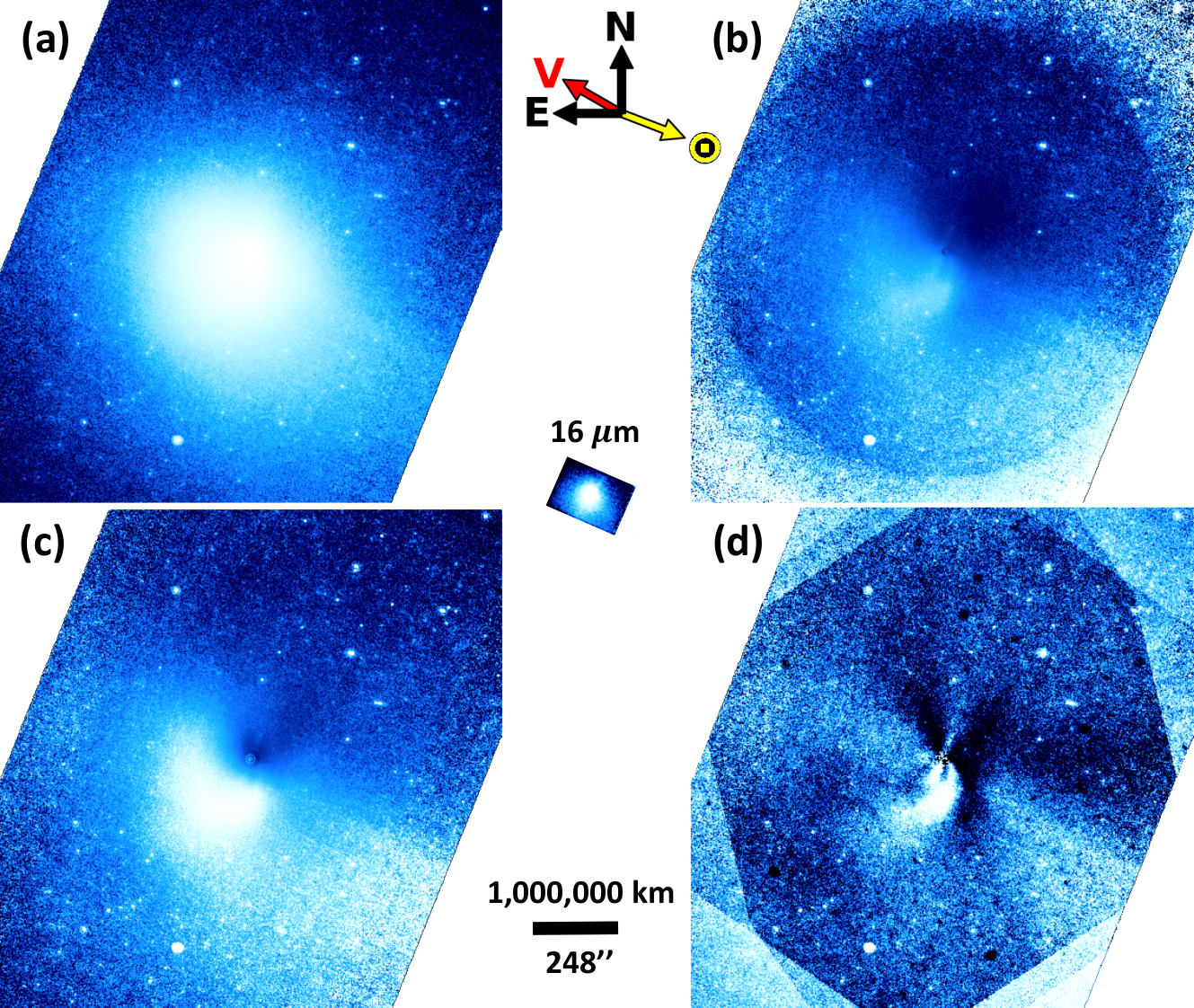}{0.8\textwidth}{}
          }
\caption{Shown is a cropped version of the 24 $\mu$m image (a), along with enhanced images: (b) division by an azimuthal average, (c) 1/$\rho$ profile removal, and (d) rotational shift differencing of 18$^{\circ}$. Equatorial north and east are indicated. The sky-plane projected directions for the Sun and SW1's heliocentric velocity vector are indicated by the yellow and red arrows. The 16 $\mu$m image (Figure \ref{fig:images}(a)) is shown to highlight the differences in the field of views between the 16 $\mu$m and 24 $\mu$m images. The large scale coma morphology shows an increased brightness in the south-west direction, possibly indicating preferential sunward emission. Also present are a more compact curved feature initially directed towards the south-southwest, curving towards the south-east and a linear feature from 1 o'clock to 7 o'clock similar to that in the 16 $\mu$m image. \label{fig:24um_images}}
\end{figure}

\begin{figure}[h!]
\gridline{
		\fig{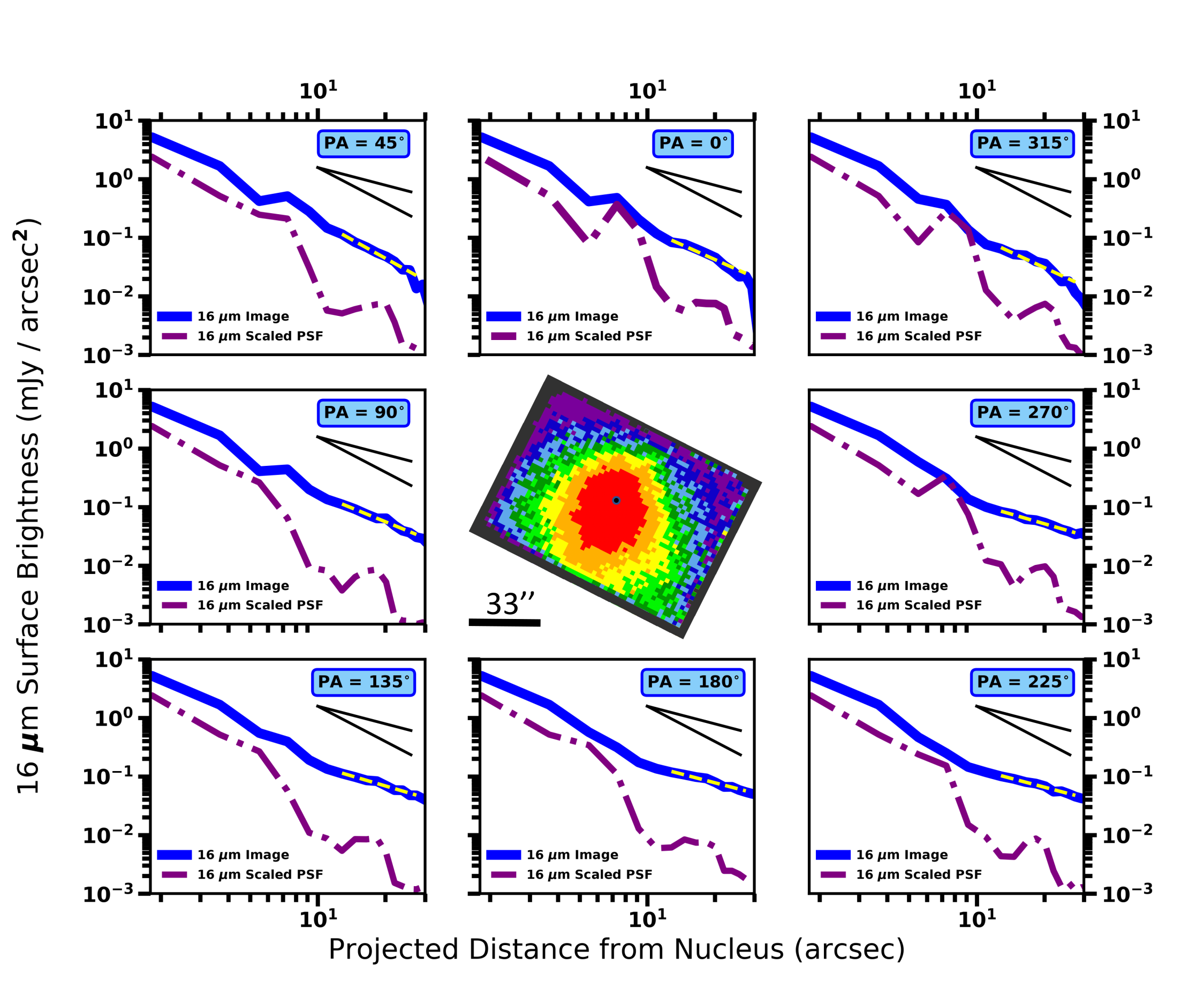}{0.9\textwidth}{}
          }
\caption{Radial profiles of the 16 $\mu$m image for different position angles in SW1. The 16 $\mu$m image is shown in the center of the plots for reference with the location of the nucleus indicated by a black circle; orientation of the image is equatorial north up and east to the left. Best-fit profiles are indicated by the yellow dashed lines for each PA and provided in Table \ref{tab:profile_slopes}. For reference, included in each plot are two black lines representing a 1/$\rho$ and 1/$\rho^2$ coma behavior. The ``roller coaster" shaped profile for the PSF is the result of the Airy diffraction pattern of the space-based telescope. \label{fig:16um_profiles}}
\end{figure}

The 16$\mu$m blue PU images (Fig. \ref{fig:images}) were obtained 1.3 days before the 24 $\mu$m MIPS images (Fig. \ref{fig:24um_images}), which were obtained on UT 2003-11-24 15:05. The 16 $\mu$m image's coma did not display any clearly distinguishable large scale radial or azimuthal features in either the un-enhanced or enhanced images. A slight enhancement on the south-southeast through south-west side of the coma is detected in the division by azimuthal average and the 1/$\rho$-removed enhanced images (Figure \ref{fig:images}(b) and (c)). This is further confirmed in Figure \ref{fig:16um_profiles}, which displays radial surface-brightness profiles for position angles (PA) at 45$^{\circ}$ spacings for the 16$\mu$m image. The radial profiles were generated by taking the median pixel value at a given radial position using 10$^{\circ}$ wide wedges center on the indicated PA. For comparison, each PA plot includes a radial profile for a scaled STINYTIM generated point spread function (PSF; \cite{kirst_2006_tinytim}) representing how a detection of SW1's bare nucleus would behave in the absence of a coma. A $C/\rho^n$ functional form was fit to the profiles for $\rho$ values between 14$''$ and 30$''$ for each PA (i.e., beyond any significant influence from the nucleus point source contribution), where $C$ is a scaling constant representing the peak coma flux near the nucleus and $n$ is the power index of the coma's profile. The fitted profile power law indices are listed in Table \ref{tab:profile_slopes}. Profiles for PAs spanning from the south-through-west directions, approximately centered on the projected sunward direction, have nearly canonical 1/$\rho$ coma profiles whereas profiles in the northeast have profile powers of approximately $n$ = 2. This asymmetric profile behavior is consistent with preferential emission of grains in the sunward direction (south-west).

\FloatBarrier

\begin{deluxetable*}{ c | c c c c }[h!]
\tablenum{2}
\label{tab:profile_slopes}
\tablecaption{16 $\mu$m and 24 $\mu$m Coma Profile Power Law Indices}
\tablewidth{0pt}
\tablehead{
\colhead{Position Angle} 	& \colhead{16 $\mu$m (14$''$ - 30$''$)} 	& \colhead{24 $\mu$m (14$''$ - 30$''$)} 	& \colhead{24 $\mu$m (30$''$ - 130$''$)} 	& \colhead{24 $\mu$m (200$''$ - 470$''$)} \\
\colhead{} 			& \colhead{(56,000 - 120,000 km)} 		& \colhead{(56,000 - 120,000 km)} 		& \colhead{(120,000 - 520,000 km)} 		& \colhead{(800,000 - 1,900,000 km)}}

\startdata
0$^{\circ}$		&	-1.7			&	-1.1		&	-0.8						&	-1.4			 			\\
45$^{\circ}$	&	-2.1			&	-0.9		&	-0.7						&	-1.5	 					\\
90$^{\circ}$	&	-1.6			&	-0.7		&	-0.7						&	-1.5	 					\\
135$^{\circ}$	&	-1.1			&	-0.6		&	-0.6						&	-1.7	 					\\
180$^{\circ}$	&	-1.0			&	-0.7		&	-0.6						&	-1.1	 					\\
225$^{\circ}$	&	-1.0			&	-0.6		&	-0.8						&	-0.8$^{\textrm{a}}$	 	\\
270$^{\circ}$	&	-1.0			&	-0.9		&	-0.9						&	-0.9$^{\textrm{a}}$	 	\\
315$^{\circ}$	&	-1.8			&	-0.9		&	-1.0						&	-1.0$^{\textrm{a}}$	 	\\
\enddata

\tablecomments{ $^{\textrm{a}}$ The coma surface brightness profile power index between 30$''$ - 470$''$ was best fitted to a single value indicated in the column to the left.}

\end{deluxetable*}

The overall coma morphology as seen in the unenhanced 24 $\mu$m (Fig. \ref{fig:24um_images}(a)) image similarly shows an increased brightness in the southwest direction. This is further confirmed by the division by an azimuthal average and 1/$\rho$-removed enhanced images. The rotational-shift-differenced enhanced image (Fig. \ref{fig:24um_images}(d)) contains a curved wing feature that \cite{stansberry_2004} attribute to a rotating jet and from which they derived an $\sim$ 60 day rotation period for SW1's nucleus. Taking into consideration the great similarity in the 16 $\mu$m and 24 $\mu$m image morphology taken 1.3 days apart, and the relationship between the projected nucleus-Sun vector and the curved wing's structure suggests that this feature is possibly not the result of nucleus rotation, but is instead due to solar radiation pressure effects on micron sized dust grains emitted in the sunward direction being turned back to form the dust tail in the north-east direction \citep{jian-yang_2014, farnham_2005, mueller_2013}. While the $\sim$ 60 day rotation period derived by the earlier work may in fact coincidentally be reflective of SW1 potentially possessing a long rotation period \citep{miles_2016, schambeau_2017, schambeau_2019}, we propose that this curved wing feature is not the result of a slowly rotating nucleus. Interestingly, the wing would be symmetric around the skyplane projected nucleus-Sun axis for the case of isotropic emission from a localized nucleus surface area. Instead it is asymmetric, indicating a possible preferential direction for dust lofting from this source region.

Similar asymmetric curved-shapes features have long been seen in broadband visible imaging data of SW1 while undergoing major outbursts. Accounts of these coma morphologies have been reported in the early works of \cite{jeffers_1956} and \cite{roemer_1958_sw1}. \cite{whipple_1980} presents a detailed analysis of SW1's outburst coma morphology as detected over a 50 year baseline, resulting in the descriptive term of ``ringtailed snorter" for this often seen curved shape feature. While it may at first seem appropriate to compare the outburst and quiescent coma morphologies, detailed analyses of SW1's dust coma while in both phases of activity \citep{hosek_2013, miles_2016, schambeau_2017, schambeau_2019} have provided descriptions of the underlying processes ongoing in both phases of activity and that the two are different. The morphology of the 24 $\mu$m quiescent coma's wing may resemble that of SW1's outburst coma; however, it was produced by different mechanisms (i.e., slow, sustained dust lofting with expansion velocities in the range of 10-50 m/s while quiescent \citep{jewitt_1990_sw1} vs. impulsive short lived dust emission at high velocities in the 100-300 m/s range during major outbursts \citep{2010MNRAS.409.1682T, schambeau_2017, schambeau_2019, feldman_1995}.)

The outer edge of the wing feature seen in the 24 $\mu$m image in the south-west direction (Fig. \ref{fig:24um_images}(d)) may indicate an approximate projected length for the turn-back distance of the grains from solar radiation pressure. Using a projected cometocentric distance of $\sim$ 90$''$ (352,000 km) for the turning point of the wing as the approximate turn-back distance and the \cite{mueller_2013} equation for turn-back distance due to solar radiation pressure, we estimate the dust coma's expansion velocity:
\begin{equation}
v = \Bigg[ \frac{2 \rho_g \beta g \sin{\alpha} }{(\cos{\gamma})^2 } \Bigg]^{1/2},
\end{equation}

\noindent where, $\rho_g$ is the projected sky-plane turn back distance of the dust grains, $\gamma$ is the angle between the initial direction of the dust grains and the sky-plane, $\beta$ is the ratio of radiation pressure acceleration to acceleration due to solar gravity, $\alpha$ is the solar phase angle of the observations, and $g = G M_{\odot} / R_H^2$ is the solar gravitational acceleration on the dust grains ($G$ is the gravitational constant, $M_{\odot}$ is the Sun's mass and $R_H$ is the heliocentric distance of the dust grains). We estimate a $\beta$ value based on equations from \cite{finson_probstein_1968} and \cite{fulle_2004}: 
\begin{equation}
\beta = \frac{C_{pr} Q_{pr}}{\rho_d d}
\end{equation} 

\noindent where $C_{pr}$ is a collection of constants equal to $3 E_{\odot} / (8 \pi c G M_{\odot})$, where $E_{\odot}$ is the Sun's mean radiation. The parameter $Q_{pr}$ is the scattering efficiency for radiation pressure for a dust grain of diameter $d$. \cite{burns_1979} provide a thorough description of $Q_{pr}$ and explain that a value of $Q_{pr} \approx 1$ is appropriate for the assumed $d = 24$ $\mu$m grains here. We use a value for the dust grain bulk density based on recent spacecraft visited comae in situ measurements: $\rho_d = 500$ kg/m$^3$ \citep{fulle_2016}. With these assumptions we arrive at an estimated value of $\beta = 0.096$. The exact value for $\gamma$ of the dust grains most dominantly contributing to the wing feature is unknown. Most probably it is the result of dust grains emitted over a continuum of angles. For this reason we calculate the outflow velocity for a range of sky-plane projected dust grain angles: $\gamma = 0^{\circ}$ ($v = 50$ m/s), $\gamma = 45.0^{\circ}$ ($v = 65$ m/s) and $\gamma = 80.0^{\circ}$ ($v = 270$ m/s).

\begin{figure}[h!]
\gridline{
		\fig{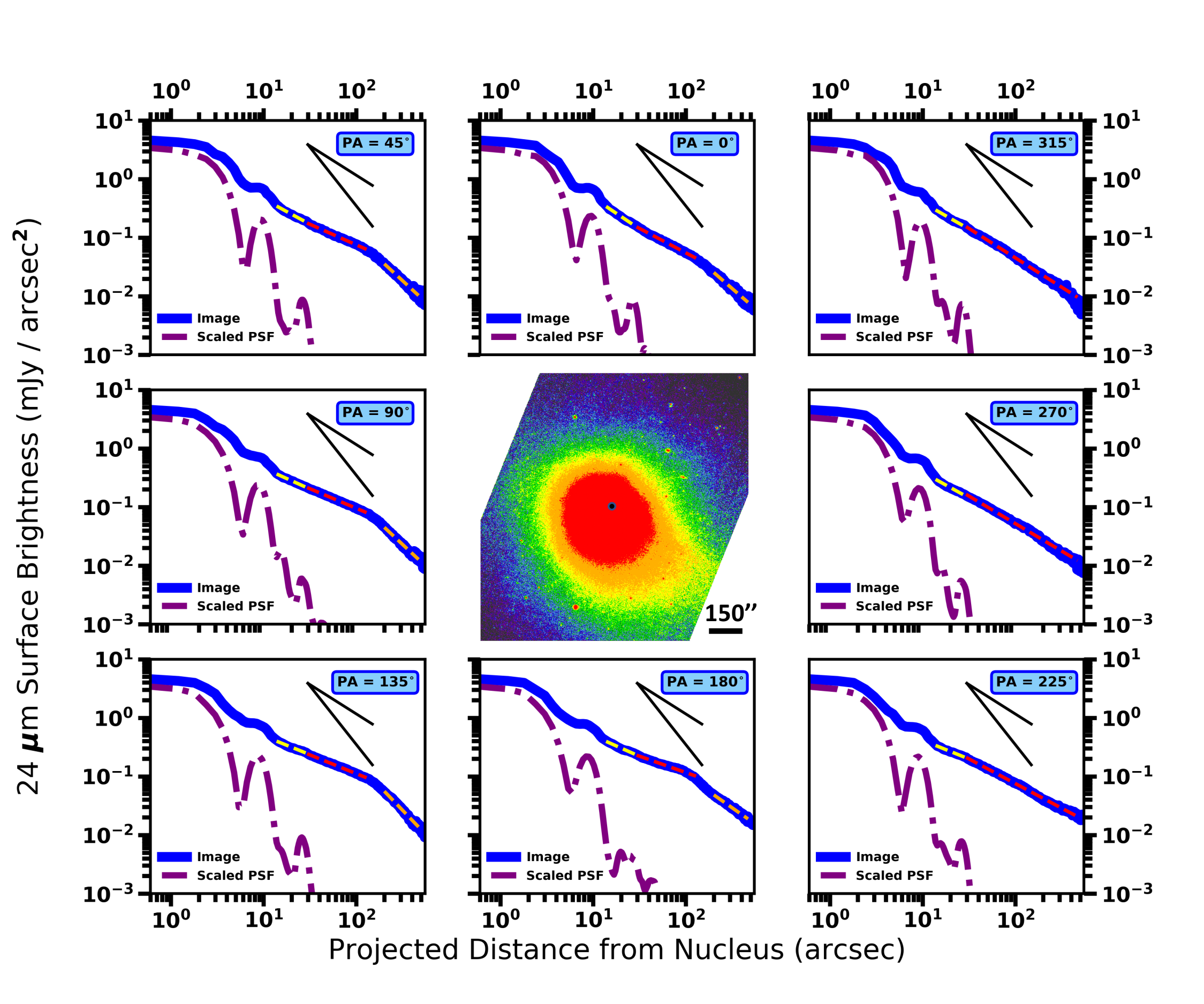}{0.9\textwidth}{}
          }
\caption{Radial profiles of the 24 $\mu$m image for different position angles. The coma morphology for radial profiles between 0$^{\circ}$ - 180$^{\circ}$ contains a knee-shaped feature at $\rho \sim$ 130$''$ (520,000 km) that is suggestive of a projected skyplane length for ongoing coma grain fragmentation and/or the projected turnback distance of dust grains from solar radiation pressure. Fitted power law indices corresponding to the yellow, red and orange curves are presented in Table \ref{tab:profile_slopes}. The location of the nucleus is indicated by the black circle in the center image. For reference, included in each plot are two black lines representing a 1/$\rho$ and 1/$\rho^2$ coma behavior. The roller coaster shaped profile for the PSF is the result of the Airy diffraction pattern of the space-based telescope. \label{fig:24um_profiles}}
\end{figure}

A similar radial surface brightness profile analysis for the 24 $\mu$m image is shown in Figure \ref{fig:24um_profiles}. The overall appearance of the coma morphology is similar to that seen in the 16 $\mu$m image, however the larger field of view (FOV) and higher S/N coma detection in the 24 $\mu$m image allows a more detailed investigation of the underlying processing ongoing within the dust coma. A change in slope of the profiles at a cometocentric distance of $\sim$ 130$''$ for PAs between 0 - 180$^{\circ}$ is suggestive of possible ongoing fragmentation for larger grains out to a projected cometocentric distance of 520,000 km (i.e., $\sim$ 130$''$). This view is supported by the coma profile's power law index being shallower than $-1$ interior to 520,000 km, suggesting an overabundance of dust grains interior to this projected distance when compared to a canonical steady-state dust emission. This behavior is possibly explained by a process of larger grains emitted from the nucleus and their subsequent fragmentation as they expand in the coma, or possibly from the decreasing size via sublimation of larger icy grains losing their volatile content. In Section \ref{sec:color_map} we discuss the possibility of icy grains in more detail. These larger (0.1 - 1.0 mm) grain populations would not contribute significantly to the 24 $\mu$m coma cross section close to the nucleus because of its relative lack of surface area, but could still easily support the observed number density of 24 micron sized grains due to a fragmentation cascade (N.B. - as long as there are particles $>>$ 24 $\mu$m in radius, they can always fragment/disrupt into many smaller particles and keep the observed particle size distribution (PSD) going) and thus maintain the coma's enhanced 24 $\mu$m surface brightness.

Similar to the 16 $\mu$m image, the coma's profiles in 24 $\mu$m close to the projected sunward direction (PAs: 225$^{\circ}$, 270$^{\circ}$, and 315$^{\circ}$) all have a single profile index close to $-1$. A possible explanation for this constant surface brightness could be a preferential sunward emission of dust grains.

\FloatBarrier
\subsubsection{$\epsilon f \rho$ Measurements and Dust Production Estimates}\label{sec:efrho}
For this analysis we calculated the $\epsilon f \rho$ parameter \citep{lisse_dust_2002, kelley_2013}, an often used proxy for dust production rates using infrared emission that is analogous to the $A f \rho$ parameter for reflected dust flux in the visible \citep{ahearn_1984}. While the assumed canonical dust coma used to derive $\epsilon f \rho$ is not valid for many comets, the utility of $\epsilon f \rho$ comes from it establishing a standard procedure for estimating comae dust production rates and allowing a relative comparison between individual comets.

The expression for $\epsilon f \rho$ used is 
\begin{equation} \label{eq:efrho}
\epsilon f \rho (\lambda) = \frac{F_{th}(\lambda)}{\pi B(\lambda, T_c)} \times \frac{\Delta^2}{\rho},
\end{equation}

\noindent where $\epsilon$ is the emissivity of the dust grains at wavelength $\lambda$, $f$ is a filling factor expressing the fraction of the photometry aperture containing dust grains, $\rho$ is the linear aperture radius centered on the nucleus which is being used to measure the flux, $F_{th}(\lambda)$ is the flux measured in the photometric aperture for wavelength $\lambda$, $B(\lambda, T_c)$ is the Planck function calculated at the color temperature $T_c$ of the dust grains, and $\Delta$ is the geocentric distance during the observation.

For the 2003 epoch of {\it Spitzer} SW1 imaging, we used properties for the dust coma derived from our earlier analysis of IRS observations of SW1. This analysis indicated the coma was dominated by sub-$\mu$m to $\mu$m-sized amorphous silicate and amorphous carbon grains at a color temperature of $\sim$ 140 K \citep{schambeau_2015}. The color temperature map shown in Section \ref{sec:color_map} also indicates dust grains at similar color temperatures, but also that there is color temperature structure present in the coma complicating the interpretation of a derived $\epsilon f \rho$ based on an assumed dust coma with uniform temperature. With these understood limitations, we used Equation \ref{eq:efrho} to calculate $\epsilon f \rho$ values for each of the three bands containing extracted coma flux measurements. Additionally, we calculated $\epsilon f \rho$ values using an expression for dust coma color temperature ($T_c = 300$ K/$\sqrt{(R_H)}$ = 125 K) based on the results of the Survey of Ensemble Physical Properties of Cometary Nuclei (SEPPCoN) for JFCs observations by {\it Spitzer} \citep{kelley_2013} and for the case of grains at an ideal blackbody temperature ($T_{bb} = 278$ K/$\sqrt{(R_H)}$ = 117 K) for comparison. The IRS-derived and SEPPCoN-derived dust color temperatures are slightly hotter than an ideal blackbody at the same heliocentric distance. Most probably this is the result of super-heated sub-$\mu$m sized amorphous carbon grains present in the dust coma \citep{hanner_1997} and/or potentially from the many emission features present in the thermal infrared region \citep{wooden_2002, markkanen_2019}.

For $F_{th}(\lambda)$ we subtracted the nucleus' contribution to SW1's overall flux in each aperture based on the scaled PSFs found during the coma removal process presented in Section \ref{sec:neatm}. Additionally, flux from background sources (some of them serendipitously detected asteroids) was removed by interpolating the dust coma behavior for regions around each background source.

\begin{figure}
\gridline{
		\fig{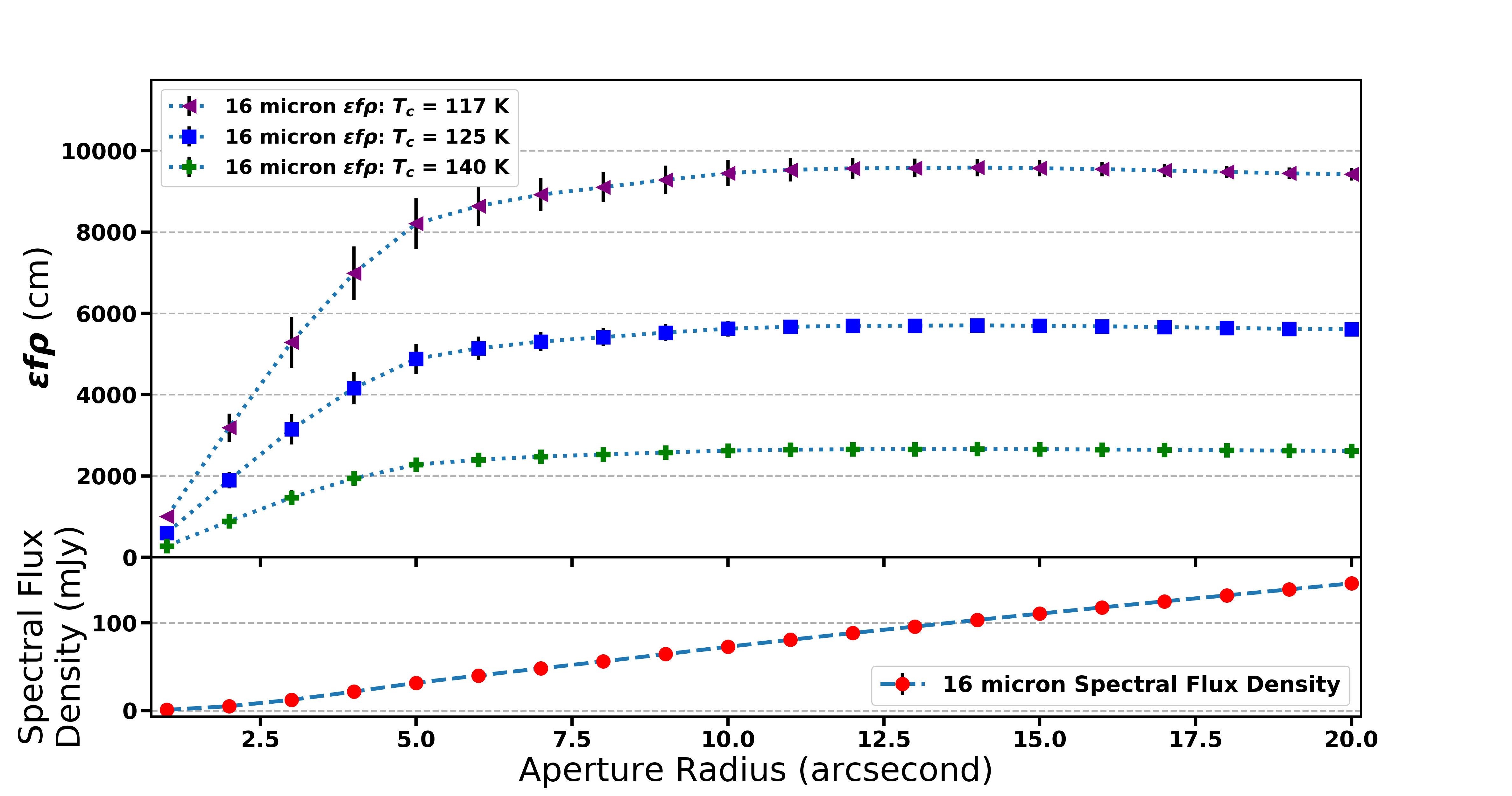}{0.95\textwidth}{}
          }
\gridline{
		\fig{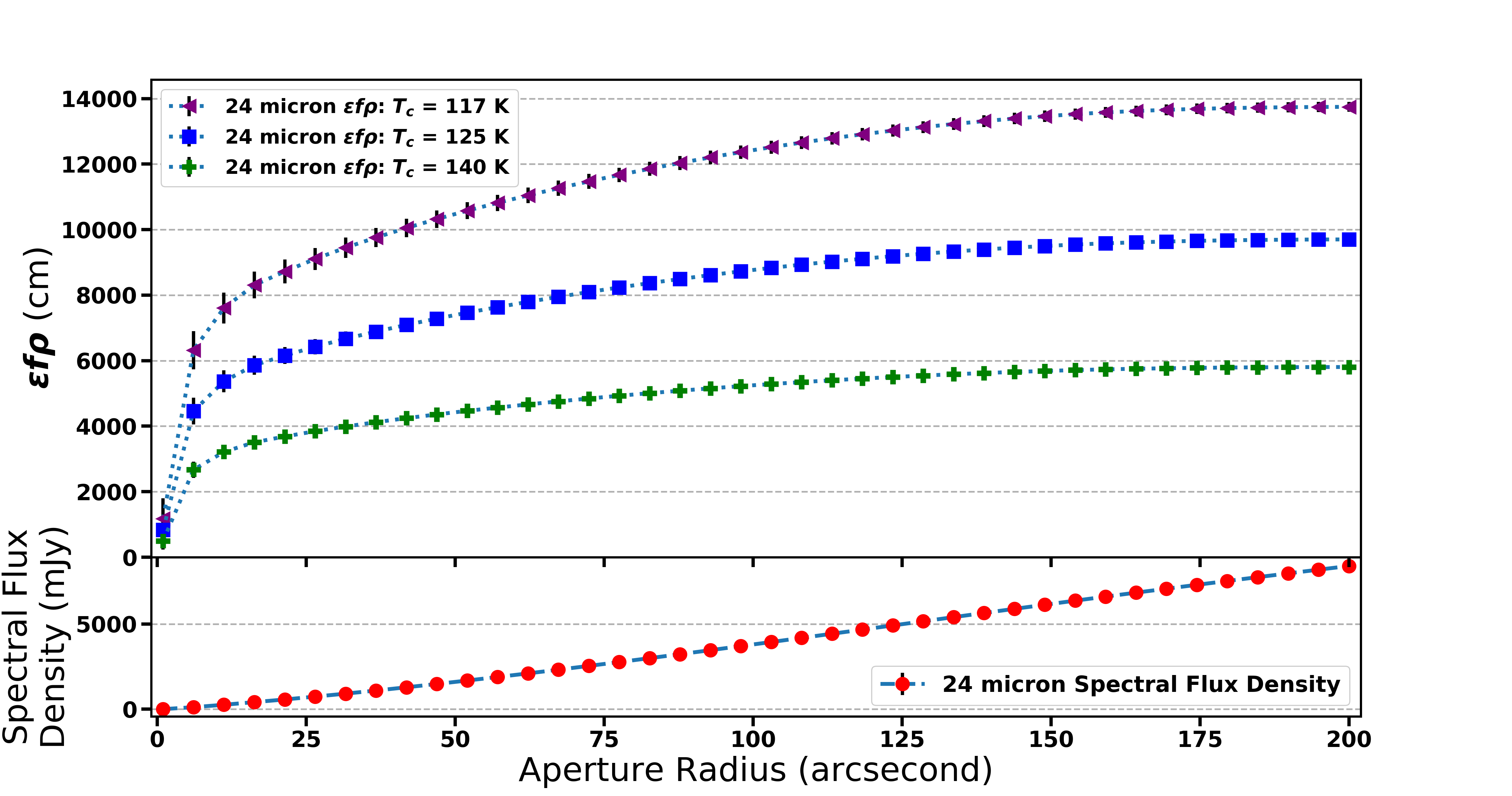}{0.95\textwidth}{}     
}
\caption{Top panel: SW1's coma spectral flux density measurements and associated $\epsilon f \rho$ measurements for the 16 $\mu$m image. Bottom panel: Similar to top, but for the 24 $\mu$m image. The 16 $\mu$m image appears to behave as a canonical dust coma with a semi-independent relationship between $\epsilon f \rho$ and aperture size, while the 24 $\mu$m has an increased value with increasing aperture size once past 5$''$. \label{fig:efrho}}
\end{figure}

Figure \ref{fig:efrho} shows plots of the 16 and 24 $\mu$m measured spectral flux density values for an array of aperture radii along with their associated $\epsilon f \rho$ measurements for the three color temperature assumptions. Table \ref{tab:dust_rates} reports the measured flux and $\epsilon f \rho$ values along with their associated uncertainties for the largest photometry apertures used for each image. The 16 $\mu$m's nearly constant $\epsilon f \rho$ value for aperture radii larger than $\sim$ 5$''$ indicates that the 3-D shape of the dust coma primarily contributing to this image maintains a nearly canonical spherical shape \citep{fink_2012}.  On the other hand, the 24 $\mu$m $\epsilon f \rho$ profile has a slight positive slope indicating deviations from a canonical 1/$\rho$ coma's expected aperture-independent constant value. The 24 $\mu$m slope behaviors support the possibility for an overabundance of 24 $\mu$m sized dust grains for larger cometocentric distances. The steep decrease for $\epsilon f \rho$ profiles for small apertures is an artifact of the coma's image being the convolution of the coma's intrinsic surface-brightness distribution with the telescope's PSF (e.g., the intrinsic surface-brightness is spread over a larger projected surface area by the convolution process resulting in a decrease in integrated flux for apertures smaller than the PSF).

To verify that the difference in aperture photometry for the coma between the 16 $\mu$m and 24 $\mu$m images is not the result of the local infrared background in each image, we compared coadded Wide-field Infrared Survey Explorer ({\it WISE}; \cite{wright_2010}) backgrounds retrieved from the W3 (12 $\mu$m) and W4 (22 $\mu$m) intensity images downloaded from the NASA/IPAC Infrared Science Archive. W3 and W4 coadded images centered on SW1's nucleus position during each epoch of imaging were compared and we found no significant differences that could explain the different photometry behaviors.
% WISE image identifiers. coadd_id (3341m061_ac51).

The 70 $\mu$m image's low S/N surface brightness coma detection did not allow a similar radial profile analysis. Instead, we report in Table \ref{tab:dust_rates} an updated $9''$ radius aperture coma flux measurement. Our earlier reported 70 $\mu$m flux density value \citep{schambeau_2015} did not included an aperture correction for the measurement, so the earlier reported flux measurement is an underestimate. Based on a new reported measurement of 103 $\pm$ 50 mJy, we calculated an $\epsilon f \rho$ value. The large uncertainty in the derived 70 $\mu$m coma flux measurement is due to the low S/N present in the mosaicked image and SW1's proximity to one of the jail-bar artifacts often present in MOPEX generated mosaicked images \citep[see][Figure 2(b)]{schambeau_2015}.

We use the measured $\epsilon f \rho$ values to estimate dust production rates during the {\it Spitzer} imaging according to:
\begin{equation}
\dot{M} = (\epsilon f \rho) \times \frac{8 a \rho_d v}{3 \epsilon},
\end{equation}  
where $a$ is the radius of the grains, $\rho_d$ is the density of the grains, and $v$ is the radial velocity of the grains lofted from the nucleus' surface. For our calculations we assumed that the diameter of the grains dominating the emitted flux for each band is equal to the effective wavelength of each band: 15.8, 23.68, and 71.42 $\mu$m. For the density of the grains we used the same value of $\rho_d$ = 500 kg/$m^3$ \citep{fulle_2016} that was used for the estimate of the dust expansion velocity. The velocity of the emitted dust grains was chosen to be 50 m/s based on the lower approximate values for dust expansion velocity from the 24 $\mu$m coma morphology and turnback distance from solar radiation pressure. While it is likely that larger grains will have slower radial velocities than smaller grains, we adopt the same value for each band, due to the observational uncertainties of the measurements. We use a value for the dust emissivity of $\epsilon=$ 0.95. Estimated dust production rates are presented in Table \ref{tab:dust_rates}.

\begin{deluxetable*}{ c c c | c c | c c | c c }
\tablenum{3}
\label{tab:dust_rates}
\tablecaption{SW1 Thermal Infrared Dust Coma Measurements}
\tablewidth{0pt}
\tablehead{
\colhead{Band}  	& 
\colhead{$\rho^{\textrm{a}}$} 	& 
\colhead{Flux} 	    & 
\colhead{ $\epsilon f \rho$ } 	&
\colhead{$\dot{M}$  }   & 
\colhead{ $\epsilon f \rho$ } 	&
\colhead{$\dot{M}$  }   & 
\colhead{ $\epsilon f \rho$ } 	&
\colhead{$\dot{M}$  }		\\
\colhead{($\mu$m)}	& 
\colhead{($''$)} 	& 
\colhead{(mJy)} 	& 
\colhead{(cm)} 		&
\colhead{(kg/s)}      & 
\colhead{(cm)} 		&
\colhead{(kg/s)}      & 
\colhead{(cm)} 		&
\colhead{(kg/s)}
%\colhead{}			& 
%\colhead{}		& 
%\colhead{}		& 
%\colhead{$T_c$ = 117 K} 				& 
%\colhead{$T_c$ = 125 K}				& 
%\colhead{$T_c$ = 140 K}							
}
\startdata
    &       &                 &       
\multicolumn{2}{c|}{$T_c$ = 117 K} &       
\multicolumn{2}{c|}{$T_c$ = 125 K} &       
\multicolumn{2}{c}{$T_c$ = 140 K} \\
\hline 
16	& 	20	& 	145 $\pm$ 2   & 9400 $\pm$ 150  & 104 $\pm$ 2	& 5600 $\pm$ 90  & 124 $\pm$ 2 		& 2600 $\pm$ 43  & 28.8 $\pm$ 0.5	\\
24	& 	20 	&	570 $\pm$ 24  & 8700 $\pm$ 360  & 144 $\pm$ 6	& 6100 $\pm$ 260 & 101 $\pm$ 4 		& 3700 $\pm$ 150 & 61 $\pm$ 3 	\\
24	& 	200 &	8403 $\pm$ 90 & 13700 $\pm$ 150 & 227 $\pm$ 3	& 9700 $\pm$ 105 & 322 $\pm$ 4 		& 5800 $\pm$ 63  & 96 $\pm$ 1 	\\
70	& 	9 	&	102 $\pm$ 50   & 2600 $\pm$ 1300 & 130 $\pm$ 65	& 2300 $\pm$ 1100& 113 $\pm$ 55		& 1800 $\pm$ 900  & 90 $\pm$ 45	\\
\enddata

\tablecomments{ $^{\textrm{a}}$ The radius of the sky-plane projected photometry aperture.}

\end{deluxetable*} 

\begin{figure}

\gridline{
		\fig{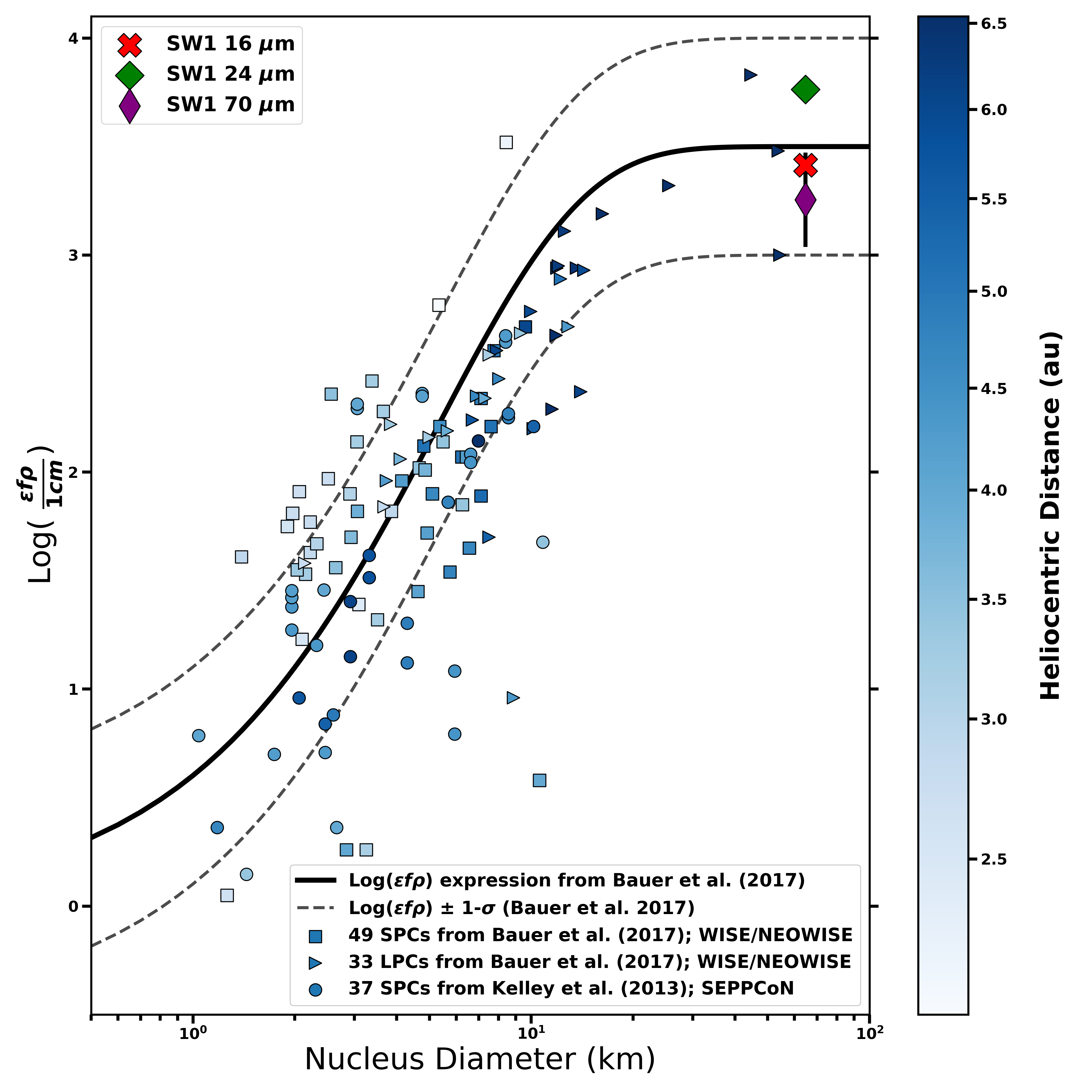}{0.75\textwidth}{}
          }

\caption{Comparison of measured $\epsilon f \rho$ values vs. nucleus diameter for comets and Centaurs from two infrared surveys and data presented here for SW1. The solid black curve indicates the empirically derived relation between $\epsilon f \rho$ vs. nucleus diameter presented in \cite{bauer_2017} using the WISE/NEOWISE detected comets; Equation 5 in this paper. Points for SW1 are based on the values from Table \ref{tab:dust_rates} for the dust temperature of $T = 140$ K. Uncertainties for the 16 $\mu$m and 24 $\mu$m points are smaller than data markers. The colors of individual markers of the WISE/NEOWISE and SEPPCoN values indicate the comet's heliocentric distance at the time of the $\epsilon f \rho$ measurement. A color bar to the right of the figure indicates the heliocentric distance color scale.}

\label{fig:efrho_panel}
\end{figure}

We have collected similar $\epsilon f \rho$ measurements based on the SEPPCoN \citep{fernandez_2013, kelley_2013} and WISE/NEOWISE \citep{bauer_2017, bauer_data_2017} surveys for comets in order to compare SW1's measured values. Results from the WISE/NEOWISE survey \citep{bauer_2017, bauer_data_2017} enabled them to develop an empirical expression relating an expected thermal dust activity for an individual comet based on its nucleus size: 
\begin{equation}
\log \Bigg( \frac{\epsilon f \rho}{\textrm{1 cm}} \Bigg) = 3.5 \Bigg( 1 - \textrm{exp}\Bigg( - \frac{D_N}{\textrm{5.3 km}} \Bigg) \Bigg) + N(0, 0.25),
\end{equation}

\noindent where $D_N$ is the nucleus diameter in km and $N(0, 0.25)$ is a Gaussian distribution with mean of 0 and variance of 0.25. In Figure \ref{fig:efrho_panel}, we plotted measurements from both infrared surveys, the empirical expression developed by \cite{bauer_2017}, and SW1's measurements from this work. 

As Figure 6 shows, Equation 5 fits the SEPCCoN $\epsilon f \rho$ values and SW1 values presented in this work. Interestingly, the expression implies that comets with nuclei diameters larger $\sim$ 20 km have a flattening of activity levels when compared to the steep increase of dust activity vs. diameter for comets between 1 km to 10 km diameters. This may be partly due to an observational bias in favor of detecting larger nuclei at larger heliocentric distances in combination to the distant activity being driven by a process other than water ice sublimation.  This comparison between SEPPCoN, NEOWISE and SW1 values is new, and the good fit of Equation 5 to the observations indicates that the equation is a robust estimator of a comet's larger grain coma activity level.

\FloatBarrier

Reports of SW1's dust production rate as derived from visible observations during periods of quiescent activity indicate a typical mass loss rate for sub-micron sized grains on the order of 1 - 50 kg/s. We arrived at these typical quiescent dust production rates using reported $A f \rho$ measurements from \cite{2010MNRAS.409.1682T} and \cite{hosek_2013}, but here we use a value of grain density $\rho_d$ = 500 kg/m$^3$ in order to be consistent with our $\epsilon f \rho$ derived dust production rates. We note that these dust production rates are upper limits due to their calculated $A f \rho$ values containing nucleus flux contributions. When compared to the estimated dust production rates as derived from the {\it Spitzer} data, which have nucleus flux contributions removed, the estimated dust production rates for grains in the range of 16 $\mu$m to 70 $\mu$m have a higher mass loss rate (Table \ref{tab:dust_rates}) than the sub-micron sized coma ($<$ 1 $\mu$m grains). It would be interesting to see if this trend of higher mass loss rate for the tens of micron sized grains is also seen during periods of major dust coma outburst (i.e., is the bulk of SW1's outburst mass loss coming from grains that are on the order of 10s of microns to 100 microns or from sub-micron sized grains), enabling investigations of the quiescent vs. outburst comae activity mechanisms.

\FloatBarrier
\subsubsection{Coma Modeling}

Another approach to determine the dust production rate is to model the thermal emission of an ensemble of particles defined by its size distribution. We used the model described in \cite{bockelee_2017} which computes the wavelength-dependent absorption coefficient and temperature of dust particles as a function of grain size using the Mie theory combined with an effective medium theory in order to consider mixtures of different materials. Effective medium theories (EMT) allow us to calculate an effective refractive index for a medium made of a matrix with inclusions of another material. The Maxwell-Garnett mixing rule is used in this model, and is also applied to consider the porosity of the grains, set to be 50\% at maximum \citep{bockelee_2017}. The infrared thermal spectrum is computed by summing the contributions of the individual dust particles.  The size distribution of the dust particles is described by a power-law  $n$($a$) $\propto$ $a^{-\beta}$, where $\beta$ is the size index and the particle radius takes values from $a_{\rm min}$ to $a_{\rm max}$. The dust density is taken equal to 500 kg/m$^3$. The effect of ice sublimation on the equilibrium grain temperature was not taken into account, as it has been shown that radiative cooling dominates over cooling by sublimation at far heliocentric distances \citep{beer_2006}.

We consider in this paper three different mixtures \citep[see][for the references for optical constants]{bockelee_2017}: 1) a matrix of amorphous carbon with inclusions of amorphous olivine with a Fe:Mg composition of 50:50; 2) a matrix of crystalline ice with inclusions of amorphous carbon; 3) a matrix of amorphous carbon with inclusions of crystalline ice. For mixture 1) the carbon/olivine mass ratio is 1, a value consistent with the organic mass fraction measured in comet 67P dust particles \citep{bardyn_2017}. Mixtures 2) and 3) have the same ice fraction in mass of $\sim$ 45\%, but have different optical properties.    

Other parameters set in the model are the dust maximum size $a_{\rm max}$ and the dust velocity as a function of particle size, described as varying $\propto$ $a^{-0.5}$, with a value of 60 m/s for 10-$\mu$m particles. The maximum liftable size from the surface of SW1's nucleus is estimated to be $a_{\rm max}$ = 250 $\mu$m, for a CO-driven activity  restricted to a spherical segment with half-angle of 45$^{\circ}$ and a total CO production rate of 4 $\times$ 10$^{28}$ s$^{-1}$, assuming our nucleus radius estimate of 32.3 km (Section \ref{sec:neatm}) and a nucleus density of 500 kg/m$^3$ \citep[V. Zakharov, personal communication, see][]{zakharov_2018, zahkarov_2021}. This CO outgassing description is consistent with CO millimeter observations \citep{gunnarsson_2008, wierzchos_2020, bockelee_2021}. 

The model was applied to simulate the flux density in a 9'' FOV radius at 16, 24 and 70 $\mu$m, for comparison with Spitzer data. Simulations were made for a minimum dust particle size $a_{\rm min}$ in the range 0.5--50$\mu$m and size indices in the range 2.5--4.6. These two parameters have indeed a strong influence on the dust thermal spectrum, with, e.g., a larger contribution from small particles for low $a_{\rm min}$ and high $\beta$ values resulting in a higher dust color temperature. The Spitzer constraints are flux densities in a 9'' FOV radius of 64 $\pm$ 2 mJy, 198 $\pm$ 14 mJy, 103 $\pm$ 50 mJy at 16, 24 and 70 $\mu$m, respectively. This corresponds to color temperatures of $T_{16/24}$ = 129 $\pm$ 5 K, based on the 16 \& 24 $\mu$m fluxes, and $T_{24/70}$ = 177$^{+52}_{-47}$ K based on the 24 \& 70 $\mu$m fluxes. $T_{16/24}$ and $T_{24/70}$ are consistent within 1-$\sigma$ with a value of $\sim$ 130 K, but the high central value of $T_{24/70}$ resulting from the relatively faint 70$\mu$m flux might suggest an excess of small particles poorly radiating at long wavelengths.    
 
\begin{figure}[h!]
\begin{center}
\includegraphics[width=0.49\textwidth, angle = 0]{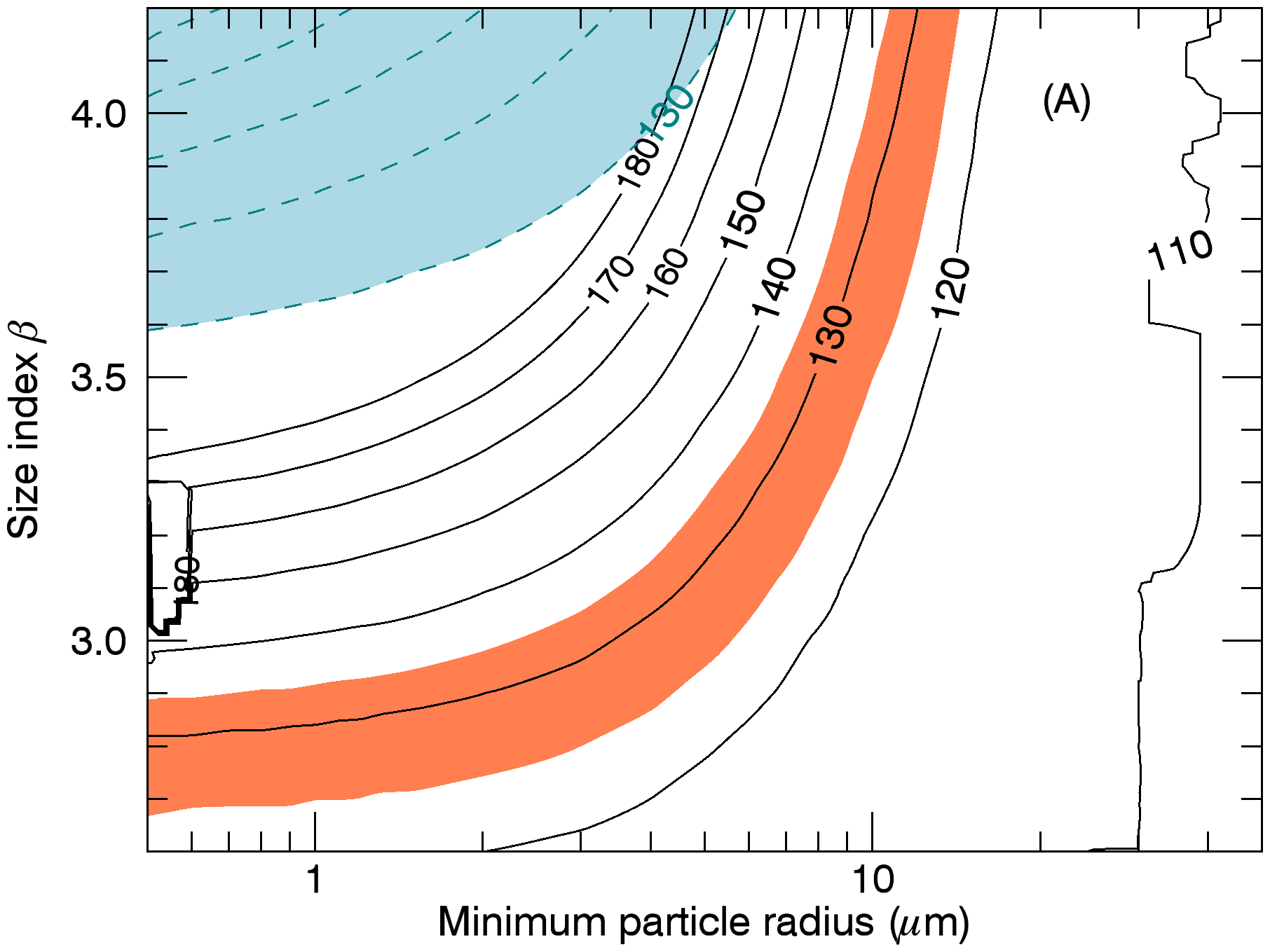}
\includegraphics[width=0.49\textwidth, angle = 0]{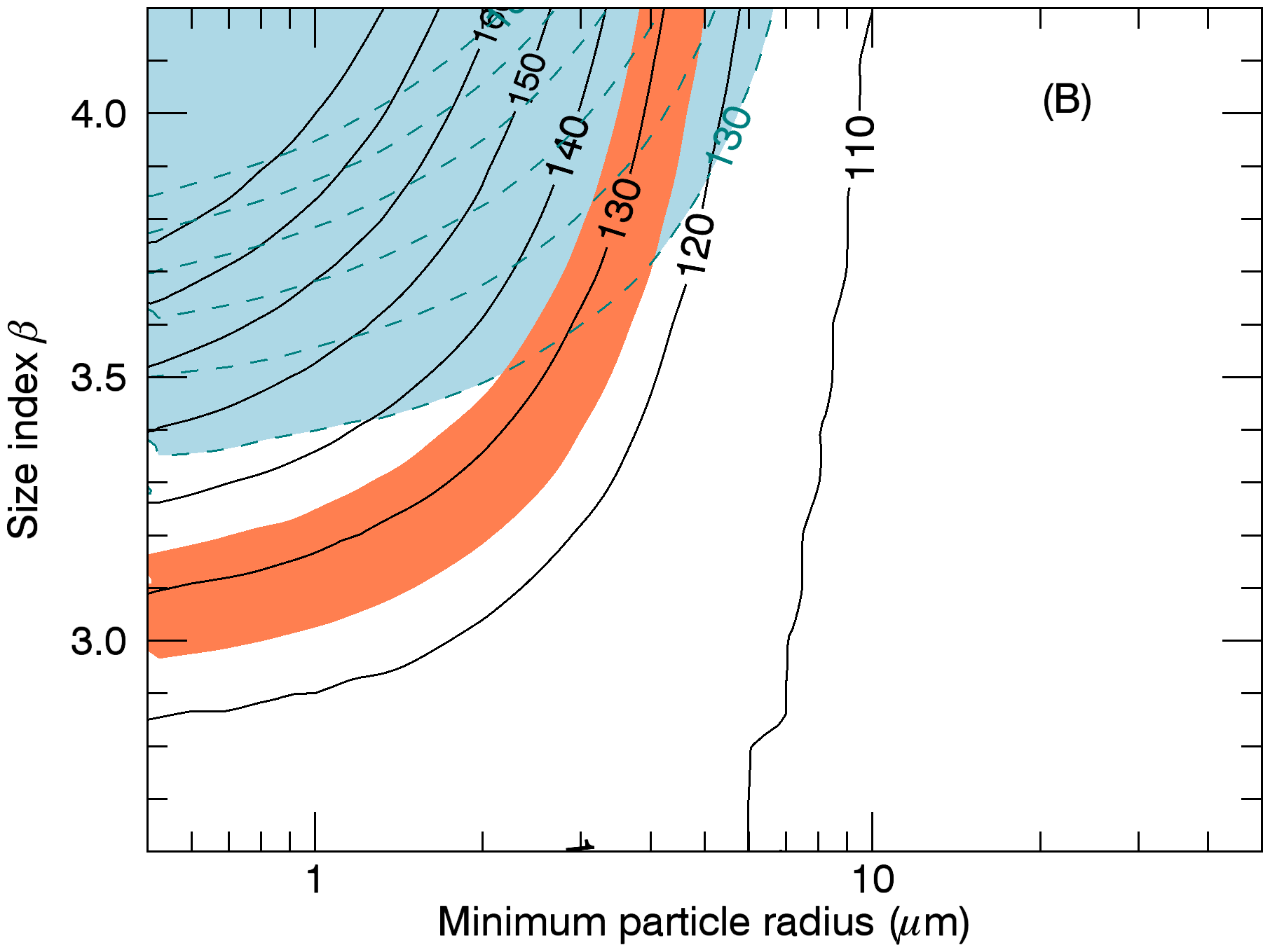}
\end{center}
 \caption{Modelled dust color temperatures $T_{16/24}$, $T_{24/70}$ as a function of minimum particle size and size index, for ice/carbon mixtures 2 (panel A) and 3 (panel B). Black plain lines show contours at constant $T_{16/24}$, in steps of 10 K. Blue dashed lines show contours at constant $T_{24/70}$, in steps of 10 K, for  $T_{24/70}$ $\geq$ 130 K. Color temperatures consistent with Spitzer measured $T_{16/24}$ and $T_{24/70}$ values are colored in orange and blue, respectively.
The assumed maximum particle size is $a_{\rm max}$ = 250 $\mu$m.}\label{fig-Tcol}
 \end{figure}

\begin{figure}[h!]
\begin{center}
\includegraphics[width=0.75\textwidth, angle = 0]{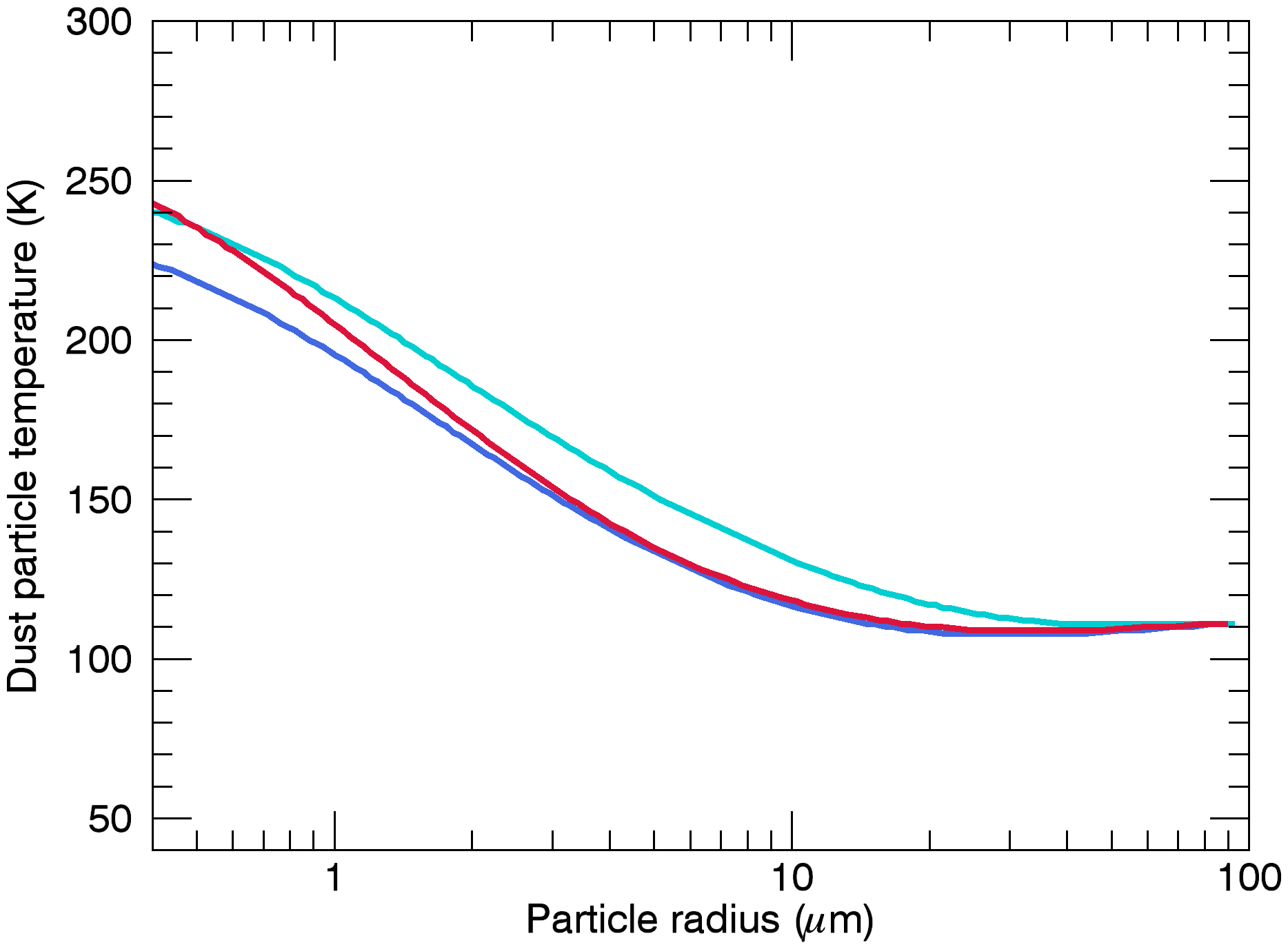} 
 \end{center}
\caption{Temperature of the dust particles as a funtion of particle radius. Results for mixtures 1 (matrix of carbon with ice inclusions), 2 (matrix of ice with carbon inclusions), and 3 (matrix of carbon with olivine inclusions) are shown in blue, turquoise and red, respectively.  }\label{fig-Tdust}
\end{figure}

Figure~\ref{fig-Tcol} shows iso-contours of $T_{16/24}$ (black plain lines) and $T_{24/70}$ (dashed blue lines) as a function of $a_{\rm min}$ and $\beta$. Domains consistent with measured  $T_{16/24}$, $T_{24/70}$ values are filled in orange and blue colors, respectively. We only show results for ice-carbon mixtures 2) and 3), since results for carbon-silicate mixture 1) are similar to those obtained for mixture 3). For mixtures 1) (not shown) and 3), the orange and blue domains overlap for $a_{\rm min}$ = 2--5 $\mu$m, whereas no overlapping is observed for mixture 2) for any set of ($a_{\rm min}$,$\beta$). Grains made of mixture 2) are hotter than other mixtures for sizes below 30$\mu$m (Fig.~\ref{fig-Tdust}), and this explains the different infrared spectra.       

\begin{figure}[h!]
\begin{center}
\includegraphics[width=0.75\textwidth, angle = 0]{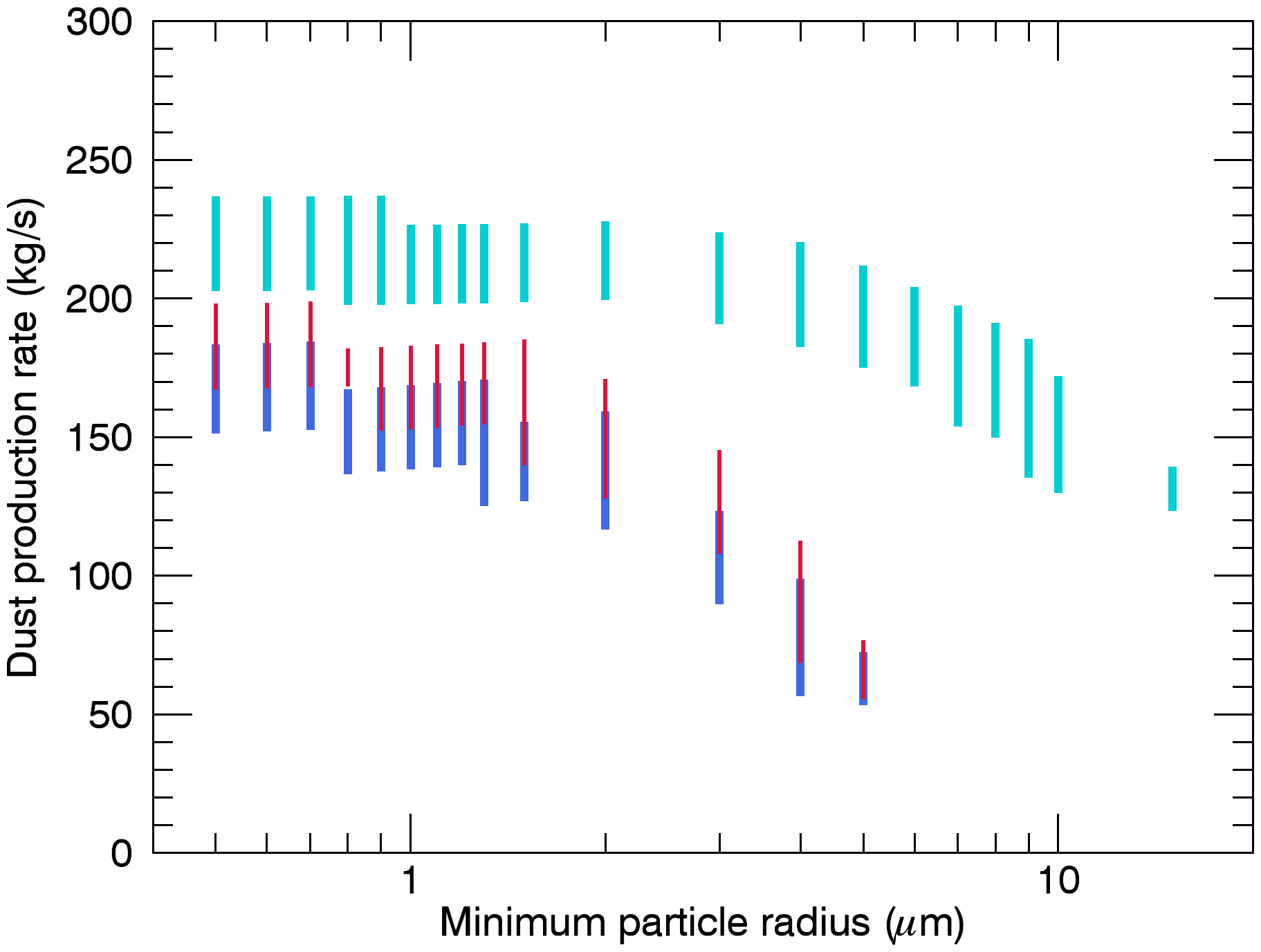}
\end{center}
\caption{Dust production rates derived from the 24-$\mu$m flux density measured in a 9'' FOV radius, using ($a_{\rm min}$,$\beta$) parameters providing color temperature $T_{16/24}$ values consistent with the measured value of 129 $\pm$ 5 K.  The range of production rate values for a given minimum size reflects the range of $\beta$ values fulfilling the requirement, and the uncertainty in the 24-$\mu$m flux. Results for mixtures 1, 2, and 3 are shown in blue, turquoise and red, respectively.  }\label{fig-Qdust}
\end{figure}

In Figure ~\ref{fig-Qdust}, we show dust production rates derived from the 24 $\mu$m flux density
using the ($a_{\rm min}$,$\beta$) parameters that provide $T_{16/24}$ values consistent with the measured value, i.e., those defining the orange region in Fig. ~\ref{fig-Tcol}. For mixtures 1) and 3) with matrices of amorphous carbon, the range is 50--200 kg/s. The low end is obtained for the highest ($a_{\rm min}$,$\beta$) values (= (5$\mu$m, 4.1--4.4)), that is a steep size distribution where 5--10 $\mu$m grains dominate the infrared emission. For size distributions with $a_{\rm min}$= 4--5 $\mu$m, the dust production rates deduced from the 24 and 70-$\mu$m fluxes are consistent, and in the range 50--100 kg/s. However, this is not the case for size distributions with small $a_{\rm min}$ values (and consequently low $\beta$ values, Fig.~\ref{fig-Tcol}), for which 70-$\mu$m derived dust production rates are by a factor 2--3 lower than those deduced from the  24-$\mu$m flux. For the ice/carbon mixture 2 (matrix of crystalline ice), the dust production rate inferred from the 24 $\mu$m flux is between 130--240 kg/s (Fig.~\ref{fig-Qdust}). The values derived from the 70 $\mu$m flux are more than 2 times lower for all sets of ($a_{\rm min}$,$\beta$) parameters. This is an expected result since for this composition, the model fails in reproducing both the $T_{16/24}$ and $T_{24/70}$ values.

The dust production rates derived with model parameters leading to a satisfactory fit to data (50--100 kg/s) are in overall agreement with those estimated in Section \ref{sec:efrho} using a simple approach. The Mie-scattering model shows that measuring dust fluxes at several wavelengths in the thermal IR can provide constraints on the particle size distribution and thermal properties. The obtained results are here limited due to the low SNR of the 70 $\mu$m dust coma flux. A flaw in the present analysis is also the known limitations of the Mie-scattering theory and of the Maxwell-Garnett mixing rule for modelling dust spectra \citep{lien_1990, mishchenko_2008}.

\FloatBarrier
\subsubsection{Coma Color Temperature Map} \label{sec:color_map}
\begin{figure}[h!]
\gridline{
		\fig{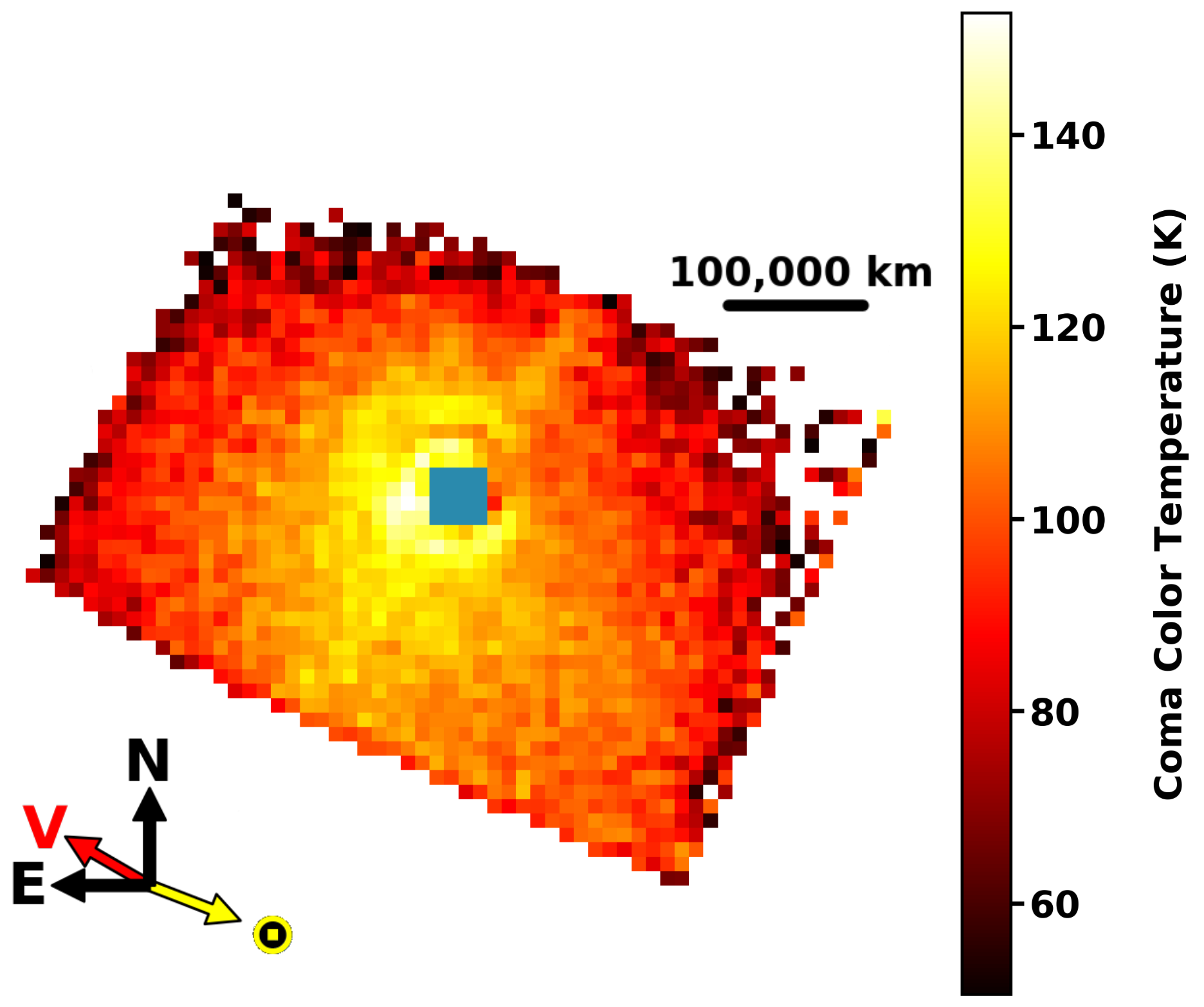}{0.75\textwidth}{}
          }
\caption{Coma color temperature map based on the 16 $\mu$m and 24 $\mu$m images.  \label{fig:temp_map}}
\end{figure}

In Figure \ref{fig:temp_map}, a color temperature map of the coma based on the 16 $\mu$m and 24 $\mu$m images is shown. This was generated by using the spectral flux density values of the coma after removal of flux contributions from the nucleus; the procedure of nucleus vs. coma flux contributions is described in Section \ref{sec:neatm} for the 16 $\mu$m image, and in our earlier work \citep{schambeau_2015} for the 24 $\mu$m data. Masked pixels identified by the teal square near the center represent regions where the PSF's subtraction may have resulted in a significant over or under subtraction for individual pixels. The white pixels on the top-left and top-right of the color map are not ``hot", but instead are masked as white due to the low S/N 16 $\mu$m detections resulting in negative spectral flux density pixel values after background subtraction. These pixels have been excluded from the color temperature fitting procedure. We note that the actual temperatures of the grains most probably are different than the values derived from fitting a Planck blackbody profile to the individual pixel values from the 16 $\mu$m and 24 $\mu$m images due to the silicate emission features present in the 24 $\mu$m bandpass and the dust coma PSD \citep{wooden_2002, markkanen_2019}. The peak temperature of the grains of $\sim$ 140 K close to the nucleus is in agreement with a color temperature derived from the IRS spectrum as analyzed in \cite{schambeau_2015}.

Overall, the general trend is a decreasing color temperature with increasing projected distance away from the nucleus. The eastern half of the coma has a higher temperature than the western side by $\sim$ 20 degrees. The interpretation of this behavior is uncertain based on the current {\it Spitzer} imaging data. We mention here plausible explanations for these color temperature behaviors based on properties of the dust coma. One possible explanation can be a population of relatively smaller grains on the eastern side of the coma composing the tail that are less efficient at radiating their stored thermal energy. Another possibility is that the western side of the coma has a higher abundance of sub-micron sized grains, resulting in an enhanced 24 $\mu$m emission above that of an ideal blackbody due to the silicate emission bands around 20 $\mu$m. The overall impact of this behavior would be a slightly lower color temperature for the western side of the coma. Future modeling efforts may be able to select between the combination of processes driving the observed color temperature, but are beyond the scope of this current work.

Using the color temperature as a proxy for the approximate dust grain temperatures and the results of \cite{beer_2006} indicates that for grain sizes on the order of tens of microns, as we have here for the 16 $\mu$m and 24 $\mu$m images, the grains have a dust mass fraction for water ice ($X$, where $X = 1$ for pure water ice) in the range of 25-50\%, with smaller grains having a higher ice content. We calculated the expected lifetimes for the water ice content of assumed spherical icy grains with diameters equal to 16 $\mu$m, 24 $\mu$m, and 70 $\mu$m and dust mass fractions $X_{16}$ = 0.5, $X_{24}$ = 0.40, $X_{70}$ = 0.25 \citep{mukai_1986-icy-lifetime, beer_2006, lien_1990}. The lifetimes of the water ice content of the grains is respectively: 112 days, 154 days, and 373 days. For these estimated lifetimes we have ignored the increased temperatures of grains as their sizes decrease due to the ongoing water ice sublimation, so our derived lifetimes are estimated upper limits. 

The presence of grains containing water ice has been inferred by the increased emissivity at longer wavelengths as derived from the modeling of SW1's {\it Spitzer} IRS spectrum \citep{schambeau_2015}. Additionally, SW1's H$_2$O production rates as derived from {\it Herschel}/HIFI observations indicate a non-nuclear extended source that is explained by the sublimation of an icy grain coma \citep{bockelee_2021}. The H$_2$O measured production rates based on {\it AKARI} and {\it Herschel} observations are in the range of $Q_{\textrm{H}_2\textrm{O}}$ $\sim$ 3 - 7$\times$10$^{27}$ molecules/s \citep{ootsubo_2012_CO+CO2, bockelee_2021}. These measured production rates are the same order of magnitude as what would be produced by the sublimation of icy grains if we use the dust-to-ice mass fractions as constrained from their color temperature and the dust production rates derived from $\epsilon f \rho$. As a first order estimate of the coma's $Q_{\textrm{H}_2\textrm{O}}$ due to the sublimation of icy grains, we calculated the production rate that would be produced from sublimation of the water ice content of icy grains following the dust production rates presented in Table \ref{tab:dust_rates}. Assuming that all of the water ice content for individual grains is fully sublimated we arrive at an estimate range of $Q_{\textrm{H}_2\textrm{O}} \sim$ (1 - 3)$\times10^{27}$ molecules/s, supporting the argument that the measured water production rates may be explained by a non-nuclear source of icy grains in the coma.

\FloatBarrier
\subsection{Nucleus Spectral Flux Density Measurements and a new NEATM}\label{sec:neatm}

To obtain nucleus photometry measurements from the blue PU images, the  flux from SW1's coma was modeled and removed. We used a well-established coma modeling technique \citep{lamy-toth_1995, lisse_1999, yan_phd_1999} for this procedure, where the azimuthal coma behavior is measured in regions outside of significant contribution from the nucleus' PSF in order to generate a synthetic coma model. The model coma's flux contribution is then subtracted from the observations resulting in an approximately bare-nucleus residual image. The residual image is then used to scale an STINYTIM generated PSF \citep{kirst_2006_tinytim} to represent the nucleus's total flux. The reader is referred to our previous work \citep{schambeau_2015} for a detailed description of this procedure.

The coma modeling and removal procedure was applied to each of the PU images resulting in six independent nucleus photometry measurements from six images at an effective 15.8 $\mu$m wavelength. The individual color corrected measurements are: 84.1, 85.0, 85.0, 87.5, 89.6, and 88.4 mJy, with a typical uncertainty of $\pm$ 7 mJy. The final measurement used for thermal modeling analysis was taken as the average of the individual measurements: 86 $\pm$ 2 mJy, with the stated 1-$\sigma$ uncertainty being the standard deviation of the six measurements.

\begin{figure}
\gridline{
		\fig{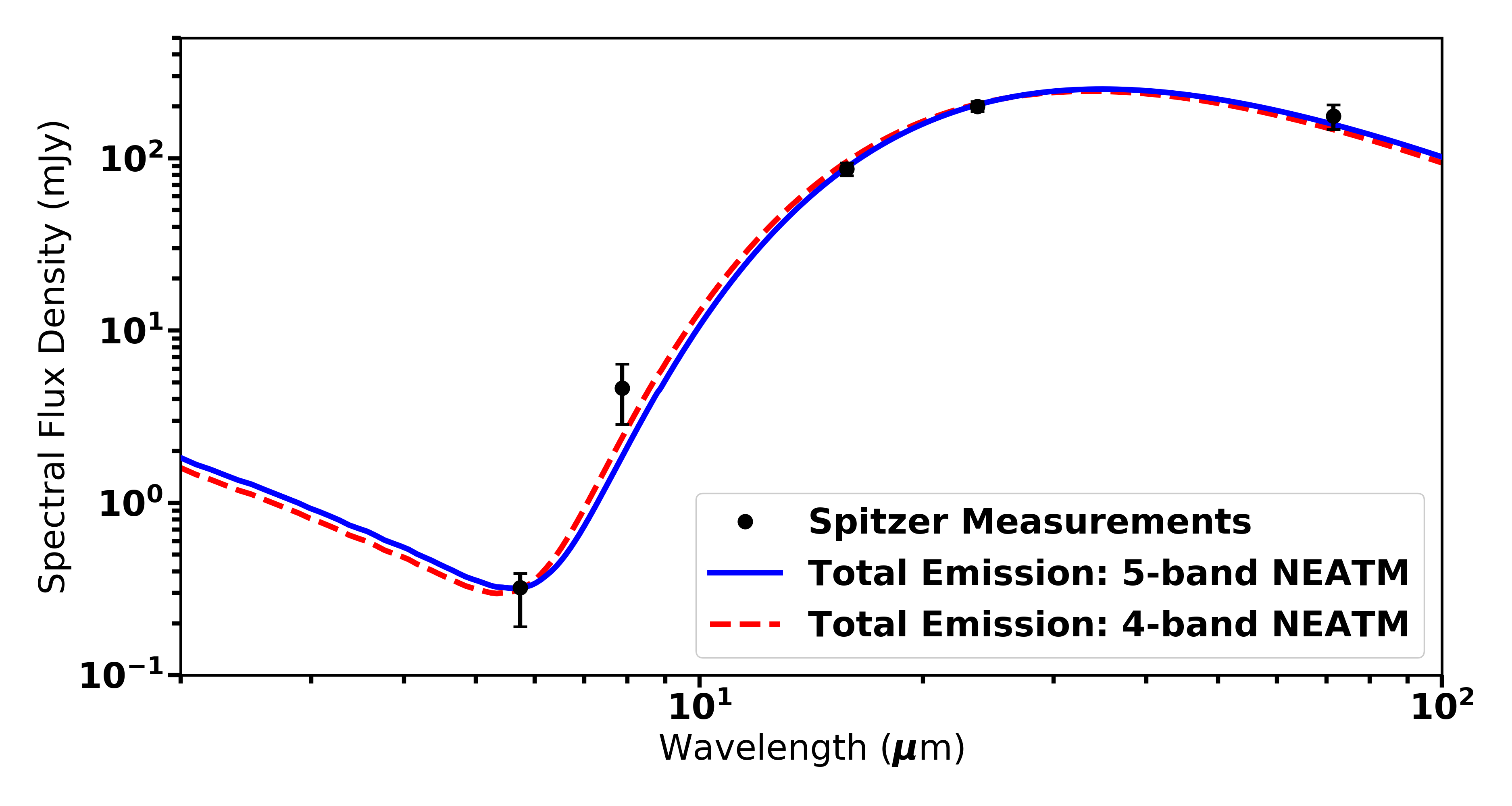}{0.9\textwidth}{}
          }
\caption{Five spectral flux density measurements of SW1,  incorporating the new blue PU data at 16$\mu$m, all acquired during 2003 November with Spitzer. Also shown is a new 5-band NEATM, which produces a nucleus radius estimate and infrared beaming parameter of $R = 32.3 \pm 3.1$ km and $\eta = 1.1 \pm 0.2$, along with the previous 4-band NEATM from \cite{schambeau_2015}. Uncertainties are 1-$\sigma$. The consistently higher than fit value for the 8 $\mu$m measurement may be the result of enhanced emission due to silicate emission bands in this region. \label{fig:neatm}}

\end{figure}

Figure \ref{fig:neatm} shows the new 15.8 $\mu$m measurement plotted along with the other four {\it Spitzer} nucleus photometry values that we reported earlier \citep{schambeau_2015}. We also plot the best-fitting 4-band thermal model (NEATM, \cite{harris_1998}) that we used in the earlier work to extract the nucleus' effective radius $R$ = $30.2^{+3.7}_{-2.9}$ km and beaming parameter $\eta$ = 0.99$^{0.26}_{-0.19}$. A re-fit using the now five spectral flux density measurements produces a nucleus size estimate and infrared beaming parameter that are slightly larger, but within the 1-sigma uncertainties of the earlier results: $R = 32.3 \pm 3.1$ km and $\eta = 1.1 \pm 0.2$. We propose these new values be used in future investigation of SW1 in lieu of our earlier analysis \citep{schambeau_2015}, because of the reduced uncertainty due to modeling with five, rather than four points. For our new NEATM analysis similar assumptions as those used for our previous work and for (e.g.) SEPPCoN \citep{fernandez_2013} were used: bolometric bond albedo $A = 0.012$ (assuming a visible-wavelength geometrical albedo $p = 0.04$ and phase integral relation $q = 0.290 + 0.684 G$, \citep{harris_lagerros_2002}, emissivity $\epsilon = 0.95$, and slope parameter $G = 0.05$.

\FloatBarrier

\section{Summary and Conclusions} \label{sec:conclusion}

A more detailed analysis of November 2003 {\it Spitzer} observations of SW1 \citep{schambeau_2015} is presented, which incorporates 16$\mu$m data for the first time, and significantly improves characterization of the Centaur's tens of microns dust coma during a period of quiescent activity.

The 16 $\mu$m blue PU images were remarkably symmetric with evidence for an $\sim$ 70 percent coma enhancement in the south-southeast direction, which may be reflective of tail formation. The 16 $\mu$m coma's morphology indicated preferential sunward emission of dust grains. No signs of grain fragmentation were indicated by the data within the image FOV (273,000 $\times$ 386,000 km).

Re-analysis of the 24 $\mu$m images reveal a large scale coma morphology of increased brightness in the southwest direction, consistent with preferential sunward emission. These data also show a more compact wing feature initially directed toward the south-southwest to a projected cometocentric distance of 352,000 km (90$''$) and curving toward the southeast. This feature has previously been interpreted to be due to the nucleus' rotation, but we propose instead that this is the result of solar radiation pressure effects and gravity on micron sized dust grains that were emitted in the sunward direction and were turned back to form a dust tail. Further analysis of this feature is encouraged. Interestingly, analysis of the 24 $\mu$m surface brightness radial profiles shows a noticeable change of slope at $\sim$ 520,000 km cometocentric distance at positions angles $\sim$ 0 through 180 degrees. This change in slope is consistent with the projected distance to the outer edge of the curved feature. We used measurements of this turning-back point of the curved feature to estimate a dust grain outflow velocity in the range of 50$-$270 m/s depending on the ejection direction of grains.

Using the improved 140 K color temperature measured from the IRS spectrum \citep{schambeau_2015} and in this work (Section \ref{sec:color_map}) we calculated the $\epsilon f \rho$ parameters: 16 $\mu$m (2600 $\pm$ 43 cm), 24 $\mu$m (5800 $\pm$ 63 cm), and 70 $\mu$m (1800 $\pm$ 900 cm). SW1's values were found to follow the $\epsilon f \rho$ vs. nucleus size relation observed from the WISE/NEOWISE observed comets \citep{bauer_2017}. Additionally, for the first time, we compare the WISE/NEOWISE and SEPPCoN \citep{kelley_2013} derived $\epsilon f \rho$ measurements and see agreement between the two surveys, strengthening the argument for the empirically derived relationship's application as a predictor of cometary comae. 

A coma model \citep{bockelee_2017} was used to constrain the coma's dust grain size distribution and mass loss rate. The model was constrained by 9$''$ radius aperture photometry measurements of 16 $\mu$m, 24 $\mu$m, and 70 $\mu$m coma flux density. Models with a dust grain composition of a matrix of amorphous carbon with inclusions of (1) amorphous olivine or (2) crystalline water ice were in agreement with the {\it Spitzer} data. The two models had similar ranges for the best-fit grain size distributions: power-law index $\beta$ ranging from 4.1 to 4.4, minimum grain size $a_{\rm min}$ ranging from 4 $\mu$m to 5 $\mu$m, and maximum grain radius $a_{\rm max}$ = 250 $\mu$m. The dust production rates derived with model parameters leading to a satisfactory fit to data (50--100 kg/s) are in overall agreement with those estimated using the measured $\epsilon f \rho$ values.

Using the 16 $\mu$m and 24 $\mu$m images we constructed a coma color-temperature map, which also peaks at $\sim$ 140 K, decreasing with increasing cometocentric distance, and an east-to-west asymmetry with the eastern coma being $\sim$ 20 degrees higher. This behavior is the result of a particle size distribution of grains of varying compositions. Future analyses of these data are encouraged to better constrain SW1's large grain coma environment.

We used the 140 K color temperature as a plausible physical temperatures for individual grains. This assumption is supported by our earlier analysis of the IRS spectrum \citep{schambeau_2015}. Using the dust production rates measured here we estimated a H$_2$O production rate from the sublimation of icy coma grains: $Q_{\textrm{H}_2\textrm{O}} \sim$ (1 - 3)$\times10^{27}$ molecules/s. This range agrees with other measurements of SW1's water production rate \citep{ootsubo_2012_CO+CO2, bockelee_2021}

Coma modeling and its removal from the IRS blue PU imaging data at 16 $\mu$m were used, along with measurements at other infrared wavelengths, to produce a nucleus radius of $R$ = 32.3 $\pm$ 3.1 km for SW1, which is within 1-$\sigma$ of and has smaller uncertainties than prior measurements using {\it Spitzer} data \citep{stansberry_2004, stansberry_2008, schambeau_2015}. This analysis also yields a slightly higher NEATM derived beaming parameter ($\eta = 1.1 \pm 0.2$). The size of SW1 places it on the smaller end of the currently-known Centaur size distribution \citep{bauer_2013, duffard_2014, lellouch_2013}, but on the larger end for small bodies with known cometary activity \citep{stansberry_2008, fernandez_2013}. With the refined nucleus size estimate presented here, we encourage future modeling efforts to better understand the bound inner coma environment of SW1. 

The Centaur SW1's large size among active objects, in combination with its orbital history that indicates it has not spent a significant amount of time interior to Jupiter \citep{sarid_2019}, positions it as a high-priority target for future observational and in situ investigations to better understand moderately sized and relatively pristine planetesimals to better understand the period of thermal evolution experienced while in the gateway transition from Centaur to JFC. We encourage the community to undertake new observations of SW1 and also for any currently existing and planned new observations to be listed on the SW1 observing campaign website:  \href{https://wirtanen.astro.umd.edu/29P/29P_obs.shtml}{wirtanen.astro.umd.edu/29P/29P\_obs.shtml}. Additionally, we provide here links to the following resources emphasizing the importance of continued observations of SW1 and best practices for new observations: (1) the call for observations from \cite{womack_2020} and (2) a guide for new observations provided by the British Astronomical Association \citep{miles_2019}. 

\FloatBarrier

\acknowledgments
We would like to thank Dr. Richard Miles for his helpful discussions during the preparation of this manuscript. Additionally, we would like to thank our two anonymous reviewers, who's thorough review of the manuscript provided improvements to the presentation of the analyses.

We also thank the NASA Earth and Space Science Fellowship (NNX16AP41H) and the Center for Lunar and Asteroid Surface Science (CLASS, NNA14AB05A) for support of this work.

This work is based on observations made with the Spitzer Space Telescope, which is operated by the Jet Propulsion Laboratory, California Institute of Technology under a contract with NASA. This research made use of Tiny Tim/Spitzer, developed by John Krist for the Spitzer Science Center. The Center is managed by the California Institute of Technology under a contract with NASA. This publication makes use of data products from the Wide-field Infrared Survey Explorer, which is a joint project of the University of California, Los Angeles, and the Jet Propulsion Laboratory/California Institute of Technology, funded by the National Aeronautics and Space Administration.


%\vspace{5mm}
%\facilities{HST(STIS)}


%\software{astropy \citep{2013A&A...558A..33A}
%          }


\bibliography{refs}{}
\bibliographystyle{aasjournal}



\end{document}